\documentclass{aa}
\usepackage{natbib,graphicx}

\bibpunct{(}{)}{;}{a}{}{,} % Astronomy & Astrophysics, ApJ

\begin{document}
\title{The carrier of the ``30'' $\mu$m emission feature in evolved
  stars\thanks{based on observations obtained with ISO, an ESA project with
  instruments funded by ESA Member states (especially the PI
  countries: France, Germany, the Netherlands and the United Kingdom)
  with the participation of ISAS and NASA}}
\subtitle{A simple model using magnesium sulfide}
\authorrunning{Hony et al.}
\titlerunning{The carrier of the ``30'' $\mu$m feature}
\author{
  S. Hony\inst{1},
  L. B. F. M. Waters\inst{1,2},
  A. G. G. M. Tielens\inst{3,4}}
\institute{ 
  Astronomical Institute ``Anton Pannekoek'', Kruislaan 403, 1098 SJ
  Amsterdam, The Netherlands
  \and
  Instituut voor Sterrenkunde, K.U. Leuven, Celestijnenlaan 200B, 3001
  Heverlee, Belgium
  \and
  SRON Laboratory for Space Research Groningen, P.O. Box 800, 9700 AV
  Groningen, The Netherlands
  \and
  Kapteyn Astronomical Institute PO Box 800, 9700 AV
  Groningen, The Netherlands}

\offprints{S.Hony, \email{hony@astro.uva.nl}} 
\date{received date; accepted date} 

\abstract{We present 2$-$45 $\mu$m spectra of a large sample of
  carbon-rich evolved stars in order to study the ``30'' $\mu$m
  feature. We find the ``30'' $\mu$m feature in sources in a wide
  range of sources: low mass loss carbon stars, extreme carbon-stars,
  post-AGB objects and planetary nebulae. We extract the profiles from
  the sources by using a simple systematic approach to model the
  continuum. We find large variations in the wavelength and width of
  the extracted profiles of the ``30'' $\mu$m feature.  We modelled
  the whole range of profiles in a simple way by using magnesium
  sulfide (MgS) dust grains with a MgS grain temperature different
  from the continuum temperature. The systematic change in peak
  positions can be explained by cooling of MgS grains as the star
  evolves off the AGB. In several sources we find that a residual
  emission excess at $\sim$26 $\mu$m can also be fitted using MgS
  grains but with a different grains shape distribution. The profiles
  of the ``30'' $\mu$m feature in planetary nebulae are narrower than
  our simple MgS model predicts. We discuss the possible reasons for
  this difference.  We find a sample of warm carbon-stars with very
  cold MgS grains. We discuss possible causes for this phenomenon. We
  find no evidence for rapid destruction of MgS during the planetary
  nebula phase and conclude that the MgS may survive to be
  incorporated in the ISM.  
  }

\maketitle
\section{Introduction}
The far infrared (IR) spectra of carbon-rich evolved objects; i.e.,
carbon-rich AGB stars (C-stars), post asymptotic giant branch objects
(post-AGBs) and planetary nebulae (PNe) are typified by a broad
emission feature around 30 $\mu$m. This ``30'' $\mu$m feature was
first discovered in the far-IR spectra of CW~Leo, IC~418 and NGC~6572
by \citet{1981ApJ...248..195F}.  Since then this feature has been
detected in a wide range of carbon-rich evolved objects from
intermediate mass loss C-stars \citep{1998Ap&SS.255..351Y} to
post-AGBs and PNe \citep{1993ais..conf...87O, 1993ais..conf..163C,
  1995ApJ...454..819O, 1999A&A...344..918J, 1999A&A...345L..39S,
  2001A&A...378L..41H}.  The feature is commonly found in C-rich
post-AGBs and PNe however with varying band shapes and varying feature
to continuum ratios \citep{1985ApJ...290L..35G, 2000tesa.conf....3W,
  2000ApJ...535..275H, Volk2002}

\citet{1985ApJ...290L..35G} proposed solid magnesium sulfide (MgS) as
the possible carrier of the ``30'' $\mu$m feature. Their suggestion is
based on the coincidence of the emission feature with the sole
IR-resonance of MgS \citep{1985ApJ...290L..41N, 1994ApJ...423L..71B}
and the fact that MgS is one of the expected condensates around these
objects \citep{1978ApJ...219..230L, 1999IAUS..191..279L}. Several
authors have taken up on this suggestion and compared observations
with laboratory measurements of MgS. These comparisons were further
facilitated by the publication of the optical constants of MgS in the
IR range by \citet{1994ApJ...423L..71B}. These authors found that the
far IR excess of CW~Leo can be successfully modelled using MgS grains
with a broad shape distribution.

More recently, \citet{1999A&A...344..918J} and
\citet{1999A&A...345L..39S} have modelled the spectra taken with the
Short Wavelength Spectrometer (SWS) \citep{1996A&A...315L..49D}
on-board the Infrared Space Observatory (ISO)
\citep{1996A&A...315L..27K} of the C-star IRAS~03313+6058 and the
post-AGB object IRAS~04296+3429 respectively.  They find that for
these sources which show a strong ``30'' $\mu$m feature, the elemental
abundances of Mg and S are consistent with MgS as the carrier of the
feature.

\citet{2000ApJ...535..275H} and \citet{Volk2002} have analysed ISO
spectra of a sample of post-AGBs.  They find that the profile of the
``30'' $\mu$m feature varies between sources. Although these authors
state that this decomposition is not unique, they find that their
``30'' $\mu$m feature is composed of two sub features: one feature
peaking near 26 $\mu$m and an other near 30 $\mu$m. Using these two
components in varying relative amounts they are able to explain the
range of features found in their sample. Based on the discovery of
these sub features they consider the carrier(s) of the ``30'' $\mu$m
feature to be unidentified.

Other materials have also been proposed as carriers of the ``30''
$\mu$m feature. \citet{2000ApJ...528..841D} suggest that the ``30''
$\mu$m feature may be indicative of carbon-based linear molecules with
specific side groups. Such molecules have strong absorption bands
throughout the 15$-$30 $\mu$m range. \citet{2000A&A...362L...9P}
discusses the possible contribution of carbonaceous dust grains with
oxygen in the structure. Some of these materials may show IR emission
in the 20$-$30 $\mu$m range. Since the optical properties of such
grains are sensitive to the exact composition they might be able to
explain the range of features found in the C-rich evolved stars.
Recently, \citet{2001ApJ...558L.129G} have proposed hydrogenated
amorphous carbon (HAC) as a possible carrier of the ``30'' $\mu$m
feature.

The ISO mission has provided an excellent database of observations to
study the properties of the ``30'' $\mu$m feature in detail and test
the suggested identifications \emph{systematically}. The wavelength
coverage of the SWS instrument (2-45 $\mu$m) is sufficient to
determine a reliable continuum. The sensitivity of the ISO
spectrograph allows detection of relatively weak features. The resolving
power of the instrument ($\lambda$/$\Delta\lambda$ = 500-1500) makes
it feasible to study possible substructure in the ``30'' $\mu$m
feature. Thus these observations allow a study of the ``30'' $\mu$m
feature in unprecedented detail in a large sample of sources.

In this paper, we investigate the shape and strength of the ``30
$\mu$m'' in a wide range of objects from visual visible C-stars,
extreme C-stars, post-AGBs to PNe in order to further test the MgS or
other identifications and map systematic differences between the
feature in different classes of sources.

Our paper is organised as follows. In Sect.~\ref{sec:observations}, we
describe the sample and the data reduction. In
Sect.~\ref{sec:continuum}, we present the way in which we modelled the
continuum in order to extract the feature properties.  In
Sect.~\ref{sec:profiles}, we present the full range of extracted
profile shapes and peak positions of the ``30'' $\mu$m feature and we
discuss the possible ways of interpreting the observed profiles. In
Sect.~\ref{sec:mgs}, we develop a simple model using MgS for the
``30'' $\mu$m feature. In Sect.~\ref{sec:modelresult}, we present the
model results and compare them to the astronomical spectra. In
Sect.~\ref{sec:correlations}, we present a correlation study between
several feature properties and stellar parameters. Finally, in
Sect.~\ref{sec:discussion}, we discuss the implications of our model
results and the consequences for the MgS identification. In
particular, we discuss possible causes for the deviating profiles and
the possibility that MgS produced in carbon-rich evolved stars will be
incorporated in the interstellar medium (ISM).

\section{Observations}
\label{sec:observations}
\begin{table*}
\caption{
    Source list. Observational details of the sources in this study.}
\label{tab:obslog}
\begin{tabular}{l r r r r r r l}
\hline
\hline
%%Column names
\multicolumn{1}{c}{Object}&
\multicolumn{1}{c}{IRAS name}&
\multicolumn{1}{c}{Obs.$^{a}$}&
\multicolumn{1}{c}{$\alpha$}&
\multicolumn{1}{c}{$\delta$}&
\multicolumn{1}{c}{TDT$^{b}$}&
\multicolumn{1}{c}{Sp./T}&
\multicolumn{1}{c}{Obj.Type}
\\
%% Column units
\multicolumn{1}{c}{}&
\multicolumn{1}{c}{}&
\multicolumn{1}{c}{Mode}&
\multicolumn{1}{c}{(J2000)}&
\multicolumn{1}{c}{(J2000)}&
\multicolumn{1}{c}{}&
\multicolumn{1}{c}{kK}&
\multicolumn{1}{c}{}
\\
\hline
\object{NGC 40}          & 00102$+$7214 & 01(3) &  00\,13\,01.10 &  $+$72\,31\,19.09 & 30003803 & WC   & PN\\
\object{IRAS 00210+6221} & 00210$+$6213 & 01(1) &  00\,23\,51.20 &  $+$62\,38\,07.01 & 40401901 &      & C-star\\
\object{IRAS 01005+7910} & 01005$+$7910 & 01(2) &  01\,04\,45.70 &  $+$79\,26\,47.00 & 68600302 & OBe  & post-AGB\\
\object{HV Cas}          & 01080$+$5327 & 01(1) &  01\,11\,03.50 &  $+$53\,43\,40.30 & 62902503 &      & C-star\\
\object{RAFGL 190}       & 01144$+$6658 & 01(2) &  01\,17\,51.60 &  $+$67\,13\,53.90 & 68800128 &      & C-star\\
\object{R Scl}$^{\dagger}$&01246$-$3248 &       &                &                   &          &      & C-star\\
        $-$              &              & 01(2) &  01\,26\,58.10 &  $-$32\,32\,34.91 & 37801213 &      &\\
        $-$              &              & 01(2) &  01\,26\,58.05 &  $-$32\,32\,34.19 & 37801443 &      &\\
\object{IRAS Z02229+6208}&Z02229$+$6208 & 01(1) &  02\,26\,41.80 &  $+$62\,21\,22.00 & 44804704 & G0   & post-AGB\\
\object{RAFGL 341}       & 02293$+$5748 & 01(1) &  02\,33\,00.16 &  $+$58\,02\,04.99 & 80002450 &      & C-star\\
\object{IRC+50 096}      & 03229$+$4721 & 01(2) &  03\,26\,29.80 &  $+$47\,31\,47.10 & 81002351 &      & C-star\\
\object{IRAS 03313+6058} & 03313$+$6058 & 01(1) &  03\,35\,31.50 &  $+$61\,08\,51.00 & 62301907 &      & C-star\\
\object{U Cam}           & 03374$+$6229 & 01(2) &  03\,41\,48.16 &  $+$62\,38\,55.21 & 64001445 &      & C-star\\
\object{RAFGL 618}       & 04395$+$3601 & 01(3) &  04\,42\,53.30 &  $+$36\,06\,52.99 & 68800561 & B0   & PN\\
\object{W Ori}           & 05028$+$0106 & 01(3) &  05\,05\,23.70 &  $+$01\,10\,39.22 & 85801604 &      & C-star\\
\object{IC 418}          & 05251$-$1244 & 01(2) &  05\,27\,28.31 &  $-$12\,41\,48.19 & 82901301 &36$^1$& PN\\
\object{V636 Mon}        & 06226$-$0905 & 01(1) &  06\,25\,01.60 &  $-$09\,07\,16.00 & 86706617 &      & C-star\\
\object{RAFGL 940}       & 06238$+$0904 & 01(2) &  06\,26\,37.30 &  $+$09\,02\,16.01 & 87102602 &      & C-star\\
\object{IRAS 06582+1507} & 06582$+$1507 & 01(2) &  07\,01\,08.40 &  $+$15\,03\,40.00 & 71002102 &      & C-star\\
\object{HD 56126}$^{\dagger}$&07134$+$1005 &    &                &                   &          & F5   & post-AGB\\
        $-$              &              & 06    &  07\,16\,10.20 &  $+$09\,59\,48.01 & 71802201 &      &\\
        $-$              &              & 06    &  07\,16\,10.30 &  $+$09\,59\,48.01 & 72201702 &      &\\
        $-$              &              & 01(3) &  07\,16\,10.20 &  $+$09\,59\,48.01 & 72201901 &      &\\
\object{CW Leo}          & 09451$+$1330 & 06    &  09\,47\,57.27 &  $+$13\,16\,42.82 & 19900101 &      & C-star\\
\object{NGC 3918}        & 11478$-$5654 & 01(1) &  11\,50\,18.91 &  $-$57\,10\,51.10 & 29900201 &      & PN\\
\object{RU Vir}          & 12447$+$0425 & 01(2) &  12\,47\,18.43 &  $+$04\,08\,41.89 & 24601053 &      & C-star\\
\object{IRAS 13416-6243} & 13416$-$6243 & 01(3) &  13\,45\,07.61 &  $-$62\,58\,18.98 & 62803904 &      & post-AGB\\
\object{II Lup}          & 15194$-$5115 & 06    &  15\,23\,04.91 &  $-$51\,25\,59.02 & 29700401 &      & C-star\\
\object{V Crb}           & 15477$+$3943 & 06    &  15\,49\,31.21 &  $+$39\,34\,17.80 & 25502252 &      & C-star\\
\object{PN K 2-16}$^{\dagger}$&16416$-$2758 &   &                &                   &          &  WC  & PN\\
        $-$              &              & 01(1) &  16\,44\,49.10 &  $-$28\,04\,05.02 & 29302010 &      &\\
        $-$              &              & 01(2) &  16\,44\,49.10 &  $-$28\,04\,05.02 & 67501241 &      &\\
\object{IRAS 16594-4656} & 16594$-$4656 & 01(1) &  17\,03\,09.67 &  $-$47\,00\,27.90 & 45800441 &      & post-AGB\\
\object{NGC 6369}        & 17262$-$2343 & 01(1) &  17\,29\,20.80 &  $-$23\,45\,32.00 & 45601901 & WC8$^2$& PN\\
\object{IRC+20 326}      & 17297$+$1747 & 01(1) &  17\,31\,54.90 &  $+$17\,45\,20.02 & 81601210 &      & C-star\\
\object{CD-49 11554}     & 17311$-$4924 & 01(2) &  17\,35\,02.41 &  $-$49\,26\,22.31 & 10300636 & BIIIe& post-AGB\\
\object{PN HB 5}         & 17447$-$2958 & 01(3) &  17\,47\,56.11 &  $-$29\,59\,39.70 & 49400104 &      & PN\\
\object{RAFGL 5416}      & 17534$-$3030 & 01(1) &  17\,56\,36.90 &  $-$30\,30\,47.02 & 12102004 &      & C-star\\
\object{T Dra}           & 17556$+$5813 & 01(2) &  17\,56\,23.30 &  $+$58\,13\,06.38 & 34601702 &      & C-star\\
\object{RAFGL 2155}      & 18240$+$2326 & 01(1) &  18\,26\,05.69 &  $+$23\,28\,46.31 & 47100261 &      & C-star\\
\object{IRAS 18240-0244} & 18240$-$0244 & 01(1) &  18\,26\,40.00 &  $-$02\,42\,56.99 & 14900804 & WC   & PN\\
\object{IRC+00 365}      & 18398$-$0220 & 01(2) &  18\,42\,24.68 &  $-$02\,17\,25.19 & 49901342 &      & C-star\\
\object{RAFGL 2256}      & 18464$-$0656 & 01(1) &  18\,49\,10.35 &  $-$06\,53\,03.41 & 48300563 &      & C-star\\
\object{PN K 3-17}       & 18538$+$0703 & 01(2) &  18\,56\,18.05 &  $+$07\,07\,25.61 & 49900640 &      & PN\\
\object{IRC+10 401}      & 19008$+$0726 & 01(1) &  19\,03\,18.10 &  $+$07\,30\,43.99 & 87201221 &      & C-star\\
\object{IRAS 19068+0544} & 19068$+$0544 & 01(1) &  19\,09\,15.40 &  $+$05\,49\,05.99 & 47901374 &      & C-star\\
\object{NGC 6790}        & 19204$+$0124 & 01(1) &  19\,22\,57.00 &  $+$01\,30\,46.51 & 13401107 &70$^3$& PN\\
\object{RAFGL 2392}      & 19248$+$0658 & 01(1) &  19\,27\,14.40 &  $+$07\,04\,09.98 & 85800120 &      & C-star\\
\object{NGC 6826}        & 19434$+$5024 & 01(4) &  19\,44\,48.20 &  $+$50\,31\,30.00 & 27200786 &50$^4$& PN\\
\object{IRAS 19454+2920} & 19454$+$2920 & 01(1) &  19\,47\,24.25 &  $+$29\,28\,11.78 & 52601347 &      & post-AGB\\
\object{HD 187885}       & 19500$-$1709 & 01(2) &  19\,52\,52.59 &  $-$17\,01\,49.58 & 14400346 & F2   & post-AGB\\
\object{RAFGL 2477}      & 19548$+$3035 & 01(1) &  19\,56\,48.26 &  $+$30\,43\,59.20 & 56100849 &      & C-star\\
\object{IRAS 19584+2652} & 19584$+$2652 & 01(1) &  20\,00\,31.00 &  $+$27\,00\,37.01 & 52600868 &      & C-star\\
\object{IRAS 20000+3239} & 20000$+$3239 & 01(1) &  20\,01\,59.50 &  $+$32\,47\,33.00 & 18500531 & G8   & post-AGB\\
\object{V Cyg}$^{\dagger}$&20396$+$4757 &       &                &                   &          &      & C-star\\
        $-$              &              & 01(2) &  20\,41\,18.20 &  $+$48\,08\,29.00 & 42100111 &      &\\
        $-$              &              & 01(2) &  20\,41\,18.20 &  $+$48\,08\,29.00 & 42300307 &      &\\
\hline
\hline
\end{tabular}
\end{table*}
\addtocounter{table}{-1}
\begin{table*}
\caption{(continued).}
\begin{tabular}{l r r r r r r l}
\hline
\hline
%%Column names
\multicolumn{1}{c}{Object}&
\multicolumn{1}{c}{IRAS name}&
\multicolumn{1}{c}{Obs.$^{a}$}&
\multicolumn{1}{c}{$\alpha$}&
\multicolumn{1}{c}{$\delta$}&
\multicolumn{1}{c}{TDT$^{b}$}&
\multicolumn{1}{c}{Sp./T}&
\multicolumn{1}{c}{Obj.Type}
\\
%% Column units
\multicolumn{1}{c}{}&
\multicolumn{1}{c}{}&
\multicolumn{1}{c}{Mode}&
\multicolumn{1}{c}{(J2000)}&
\multicolumn{1}{c}{(J2000)}&
\multicolumn{1}{c}{}&
\multicolumn{1}{c}{kK}&
\multicolumn{1}{c}{}
\\
\hline
\object{NGC 7027}$^{\dagger}$          &&       &                &                   &          &200$^5$& PN\\
        $-$              &              & 01(4) &  21\,07\,01.71 &  $+$42\,14\,09.10 & 02401183 &      &\\
        $-$              &              & 01(1) &  21\,07\,01.70 &  $+$42\,14\,09.10 & 23001356 &      &\\
        $-$              &              & 01(2) &  21\,07\,01.70 &  $+$42\,14\,09.10 & 23001357 &      &\\
        $-$              &              & 01(3) &  21\,07\,01.70 &  $+$42\,14\,09.10 & 23001358 &      &\\
        $-$              &              & 01(4) &  21\,07\,01.63 &  $+$42\,14\,10.28 & 55800537 &      &\\
\object{S Cep}           & 21358$+$7823 & 01(1) &  21\,35\,12.80 &  $+$78\,37\,28.20 & 56200926 &      & C-star\\
\object{RAFGL 2688}      &              & 01(3) &  21\,02\,18.80 &  $+$36\,41\,37.79 & 35102563 & F5   & post-AGB\\
\object{RAFGL 2699}      & 21027$+$5309 & 01(1) &  21\,04\,14.70 &  $+$53\,21\,02.99 & 77800722 &      & C-star\\
\object{IC 5117}         & 21306$+$4422 & 01(1) &  21\,32\,30.83 &  $+$44\,35\,47.29 & 36701824 &77$^3$& PN\\
\object{RAFGL 5625}      & 21318$+$5631 & 01(1) &  21\,33\,22.30 &  $+$56\,44\,39.80 & 11101103 &      & C-star\\
\object{IRAS 21489+5301} & 21489$+$5301 & 01(1) &  21\,50\,45.00 &  $+$53\,15\,28.01 & 15901205 &      & C-star\\
\object{SAO 34504}       & 22272$+$5435 & 01(2) &  22\,29\,10.31 &  $+$54\,51\,07.20 & 26302115 &  G5  & post-AGB\\
\object{IRAS 22303+5950} & 22303$+$5950 & 01(1) &  22\,32\,12.80 &  $+$60\,06\,04.00 & 77900836 &      & C-star\\
\object{IRAS 22574+6609} & 22574$+$6609 & 01(2) &  22\,59\,18.30 &  $+$66\,25\,49.01 & 39601910 &      & post-AGB\\
\object{RAFGL 3068}      & 23166$+$1655 & 01(2) &  23\,19\,12.48 &  $+$17\,11\,33.40 & 37900867 &      & C-star\\
\object{RAFGL 3099}      & 23257$+$1038 & 01(1) &  23\,28\,16.90 &  $+$10\,54\,40.00 & 78200523 &      & C-star\\
\object{IRAS 23304+6147} & 23304$+$6147 & 01(3) &  23\,32\,44.94 &  $+$62\,03\,49.61 & 39601867 &  G2  & post-AGB\\
\object{IRAS 23321+6545} & 23321$+$6545 & 01(1) &  23\,34\,22.53 &  $+$66\,01\,50.41 & 25500248 &      & post-AGB\\
\object{IRC+40 540}      & 23320$+$4316 & 01(2) &  23\,34\,27.86 &  $+$43\,33\,00.40 & 38201557 &      & C-star\\
\hline
\multicolumn{8}{c}{non detections}\\
\hline
\object{R For}           & 02270$-$2619 & 01(1) &  02\,29\,15.30 &  $-$26\,05\,56.18 & 82001817 &      & C-star\\
\object{SS Vir}          & 12226$+$0102 & 01(1) &  12\,25\,14.40 &  $+$00\,46\,10.20 & 21100138 &      & C-star\\
\object{Y CVn}           & 12427$+$4542 & 01(2) &  12\,45\,07.80 &  $+$45\,26\,24.90 & 16000926 &      & C-star\\
\object{RY Dra}          & 12544$+$6615 & 01(3) &  12\,56\,25.70 &  $+$65\,59\,39.01 & 54300203 &      & C-star\\
\object{C* 2178}         & 14371$-$6233 & 01(1) &  14\,41\,02.50 &  $-$62\,45\,54.00 & 43600471 &      & C-star\\
\object{V1079 Sco}       & 17172$-$4020 & 01(1) &  17\,20\,46.20 &  $-$40\,23\,18.10 & 46200776 &      & C-star\\
\object{T Lyr}           & 18306$+$3657 & 06    &  18\,32\,19.99 &  $+$36\,59\,55.50 & 36100832 &      & C-star\\
\object{S Sct}           & 18476$-$0758 & 01(2) &  18\,50\,19.93 &  $-$07\,54\,26.39 & 16401849 &      & C-star\\
\object{V Aql}           & 19017$-$0545 & 01(2) &  19\,04\,24.07 &  $-$05\,41\,05.71 & 16402151 &      & C-star\\
\object{V460 Cyg}        & 21399$+$3516 & 01(1) &  21\,42\,01.10 &  $+$35\,30\,36.00 & 74500512 &      & C-star\\
\object{PQ Cep}          & 21440$+$7324 & 01(1) &  21\,44\,28.80 &  $+$73\,38\,03.01 & 42602373 &      & C-star\\
\object{TX Psc}          & 23438$+$0312 & 06    &  23\,46\,23.57 &  $+$03\,29\,13.70 & 37501937 &      & C-star\\
\hline
\hline
\end{tabular}
\\
$^{a}$ SWS observing mode used
\citep[see][]{1996A&A...315L..49D}. Numbers in brackets correspond to
the scanning speed.\\
$^{b}$TDT number which uniquely identifies each ISO observation.\\
$^{\dagger}$These spectra have been obtained by co-adding the separate
SWS spectra also listed in the table, see text.\\
Effective temperatures from $^1$\citet{1992A&A...260..329M},
$^2$\citet{1991ApJS...76..687P}, $^3$\citet{1991ApJ...372..215K},
$^4$\citet{1995AAS...187.8008Q} and $^5$\citet{2000ApJ...539..783L}.
\end{table*}

\begin{figure}
  \includegraphics[width=8.8cm]{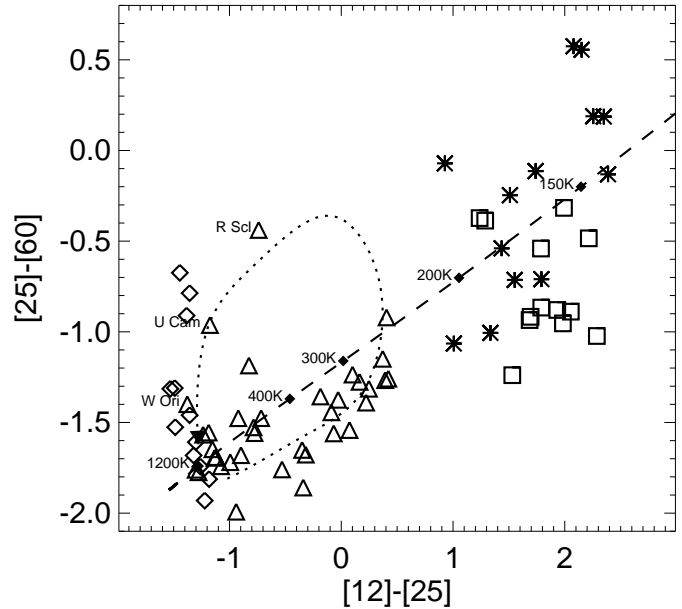}
  \caption{The IRAS two-colour diagram for the sources studied in this
    sample. The triangles represent the C-stars, the squares are the
    post-AGB objects, the stars are the PNe and the diamonds are the
    C-stars without a ``30'' $\mu$m feature detected. The dashed line
    represents the position of a blackbodies of different
    temperatures.  The dotted line sketches the evolution of a C-star
    with a detached, expanding and cooling circumstellar shell.}
  \label{fig:irascolours}
\end{figure}
We present observations obtained with ISO of a sample of bright IR
sources at different stages along the evolutionary track from C-star
via post-AGB object to PN. The observations presented here consist of
data obtained with the ISO/SWS using astronomical observing template
06 and 01 at various speeds.  These observing modes produces
observations from 2.3 to 45 $\mu$m with a resolving
power($\lambda/\Delta\lambda$) ranging from 500 to 1500.  The sample
consists of all carbon-rich evolved objects in the ISO archive which
exhibit a ``30'' $\mu$m feature stronger than 8 Jy peak intensity and
have been observed over the full 2.3$-$45.2 $\mu$m wavelength range of
the ISO/SWS. This peak intensity and the typical noise level of SWS
band 4 (29$-$45.2 $\mu$m) allows to extract a reasonably reliable
feature strength and profile. The complete wavelength coverage is
needed in order to provide a sufficient baseline to estimate the
continuum. We have further completed the sample with all observed
C-stars with an IRAS 25 $\mu$m flux over 13 Jy. These sources serve as
a control group since we would expect to detect the ``30'' $\mu$m
feature based on this brightness, the typical noise levels and the
typical feature over continuum level. These sources without the ``30''
$\mu$m feature detected are listed separately in
Table~\ref{tab:obslog}. It should be emphasised that the ISO archive
does not contain a statistically representative sample of objects. The
database of observations for the carbon stars provides a reasonable
sampling over stellar properties (e.g. mass-loss rates or colour
temperatures).  However the post-AGB sample is heavily biased towards
the ``21'' $\mu$m objects; a peculiar type of C-rich post-AGB object.
The sample of PNe contains a collection of either bright, well-known
or well-studied objects without a proper statistical selection.  It
also contains a relatively large proportion of PNe with hydrogen-poor
central stars.  The total sample of 75 sources contains 48 C-stars, 14
post-AGB objects and 13 PNe. We have detected the ``30'' $\mu$m
feature in 36 out of 48 C-stars.

We present in Fig.~\ref{fig:irascolours} the IRAS two-colour diagram
for the sources in our sample following \citet{1988A&A...194..125V}.
There are four sources in our sample without an entry in the IRAS
point source catalogue. For these sources we have used ISO/SWS and LWS
observations at 12, 25, 60 and 100 $\mu$m to calculate the IRAS
colours. For IRAS~Z02229, no measurements at 60 and 100 $\mu$m are
available.  In Fig.~\ref{fig:irascolours}, the warmest sources are
located in the lower left corner. These are the optically visible
carbon stars with a low present-day mass-loss rate
($\dot{M}\simeq10^{-8}-10^{-7}\,M_{\sun}$). With increasing mass loss
the stars become redder and move up and to the right. After the AGB,
when the mass loss has terminated, the dust moves away from the star
and cools; i.e., these sources move further to the top-right corner of
the diagram. The C-stars located above the main group of C-stars have
a clear 60 $\mu$m excess. This is evidence for an additional cool dust
component. Some of these sources are known to have an extended or
detached dust shell around them \citep{1993ApJS...86..517Y}. The empty
region between the C-stars and the post-AGBs is physical. When the
mass loss stops the star quickly loses its warmest dust and within a
short time span ($<$ 1000 yr) the star moves to the right in the
two-colour diagram. Notice how the sources without a detected ``30''
$\mu$m feature cluster on the left of the diagram, i.e., among the
warmest C-stars.

\subsection{Data reduction}
\label{sec:data}
The SWS data were processed using SWS interactive analysis product;
IA$^3$ \citep[see][]{1996A&A...315L..49D} using calibration files and
procedures equivalent with pipeline version 10.1. Further data
processing consisted of extensive bad data removal primarily
  to remove the effects of cosmic ray hits and rebinning on a fixed
resolution wavelength grid.  If a source has been observed multiple
times and these observations are of similar quality and of comparable
flux-level these data are co-added after the pipeline reduction. These
sources are indicated in Table~\ref{tab:obslog} with a dagger
($^\dagger$).  Since the features we discuss here are fully resolved
in all observing modes, we combine the data obtained in all different
modes to maximise the S/N.  Although the wavelength coverage of the
SWS instrument is well suited to study the profile of the ``30''
$\mu$m feature, there are some important instrumental effects which
hamper the unbiased extraction of the emission profiles. We discuss
these below.

\subsubsection{Splicing}
\begin{figure}
  \includegraphics[width=8.8cm]{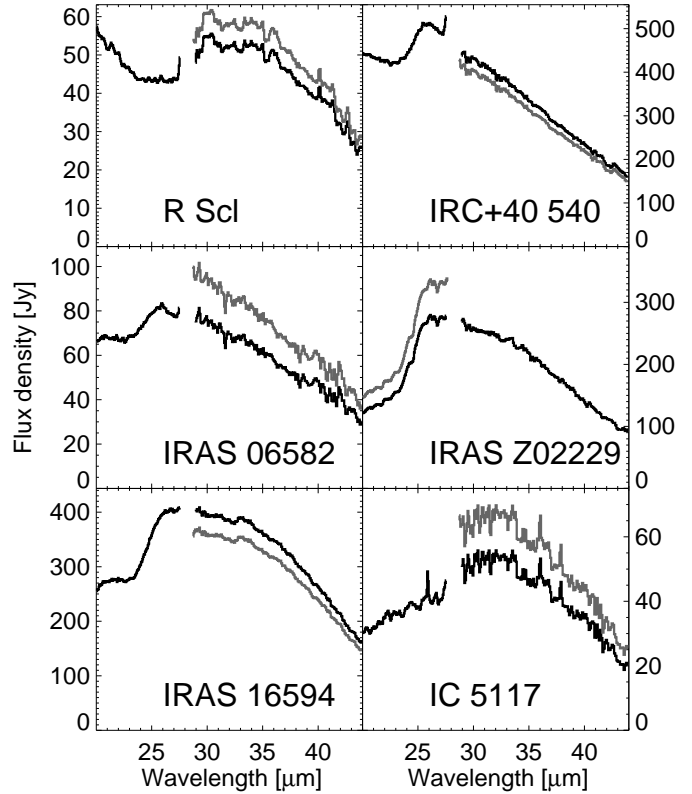}
  \caption{Examples of the splicing of the SWS band 3D (19.5$-$27.5
    $\mu$m) and 4 (28.9$-$45.2 $\mu$m) data. We show the data before
    (grey line) and after splicing (black line). All data have been
    scaled to form a continuous spectrum. As can be seen; after
    splicing, the slope of band 3D and band 4 match. We do not show
    the band 3E data. The sharp rise at 27 $\mu$m in R~Scl and
    IRC+40~450 is an instrumental artifact (see text for details).}
  \label{fig:splicing}
\end{figure}
One complete SWS AOT01 spectrum is obtained in 12 different subbands.
These subbands are observed through 3 different rectangular apertures
which range in size from 14$\arcsec\times$20$\arcsec$ at the shortest
wavelengths to 20$\arcsec\times$33$\arcsec$ at the longest
wavelengths. All these data are independently flux calibrated and need
to be combined to form one continuous spectrum for one source. We
apply scaling factors to combine the different subbands to obtain the
continuous spectra. The C-stars and post-AGB objects we present in
this study all have a small angular extent even compared to the
smallest aperture used.  Therefore we don't expect large jumps to be
present due to the differences between the apertures used. The angular
extent of some PNe can be large compared to the sizes of the
apertures. If there is a clear indication of flux jumps due to
aperture changes we have not included the source in our sample.

\subsubsection{Leakage}
At wavelengths longer than between 26 and 27.5 $\mu$m the data of SWS
subband 3D are affected by leakage adding flux from the 13 $\mu$m
region. The sources used to derive the instrumental response function
are all stellar sources without circumstellar material. These
calibration sources are all very blue and emit much more flux at 13
$\mu$m relative to 26 $\mu$m than the cool, red sources we present in
this study. Therefore these calibrators are more affected by the
leakage than our sources.  The instrumental response function derived
in this way has been implicitly corrected for leakage for the blue
sources. This resulted in fluxes in red sources to be systematically
underestimated.  More recent calibrations ($\ge$ OLP 10.0), have been
corrected for this effect.  With the improved calibrations, the
resulting slopes of the spectra beyond 26 $\mu$m have been checked and
are in general agreement with the slope of subband 4.

\subsubsection{The 27.0-27.5 and 27.5-29.0 $\mu$m region}
At wavelengths longer than 27.0 $\mu$m the data of subband 3D show a
sharp increase which is found throughout the complete database of
ISO/SWS observations independent of source type. The data of subband
3E (27.5-29.0 $\mu$m) are generally unreliable both in shape and
absolute flux level. These combined instrumental effects make it
inherently difficult to interpret the 27-29 $\mu$m spectra. Any
substructure detected solely in this region alone should be
distrusted.

The instrumental effects between 27 and 29 $\mu$m and the fact that
each of the subbands is independently flux calibrated make it
necessary to devise a strategy for splicing the band 3D, 3E and 4
data. There is unfortunately no objective way to choose this strategy.
We choose to assume minimal spectral structure between the end of
subband 3D and the beginning of band 4, i.e. to splice the subband
3D$-$4 data in such a way that the matching slopes of 3D and 4 also
match in flux level. Some examples are shown in
Fig.~\ref{fig:splicing}.  The observed discontinuities between
subbands are relatively small ($<$ 20 per cent) and can be understood
as the result of absolute flux calibration uncertainties alone.

\subsection{Full spectra}
\label{sec:spectra}
\begin{figure*}[!hp]
  \includegraphics[width=17cm]{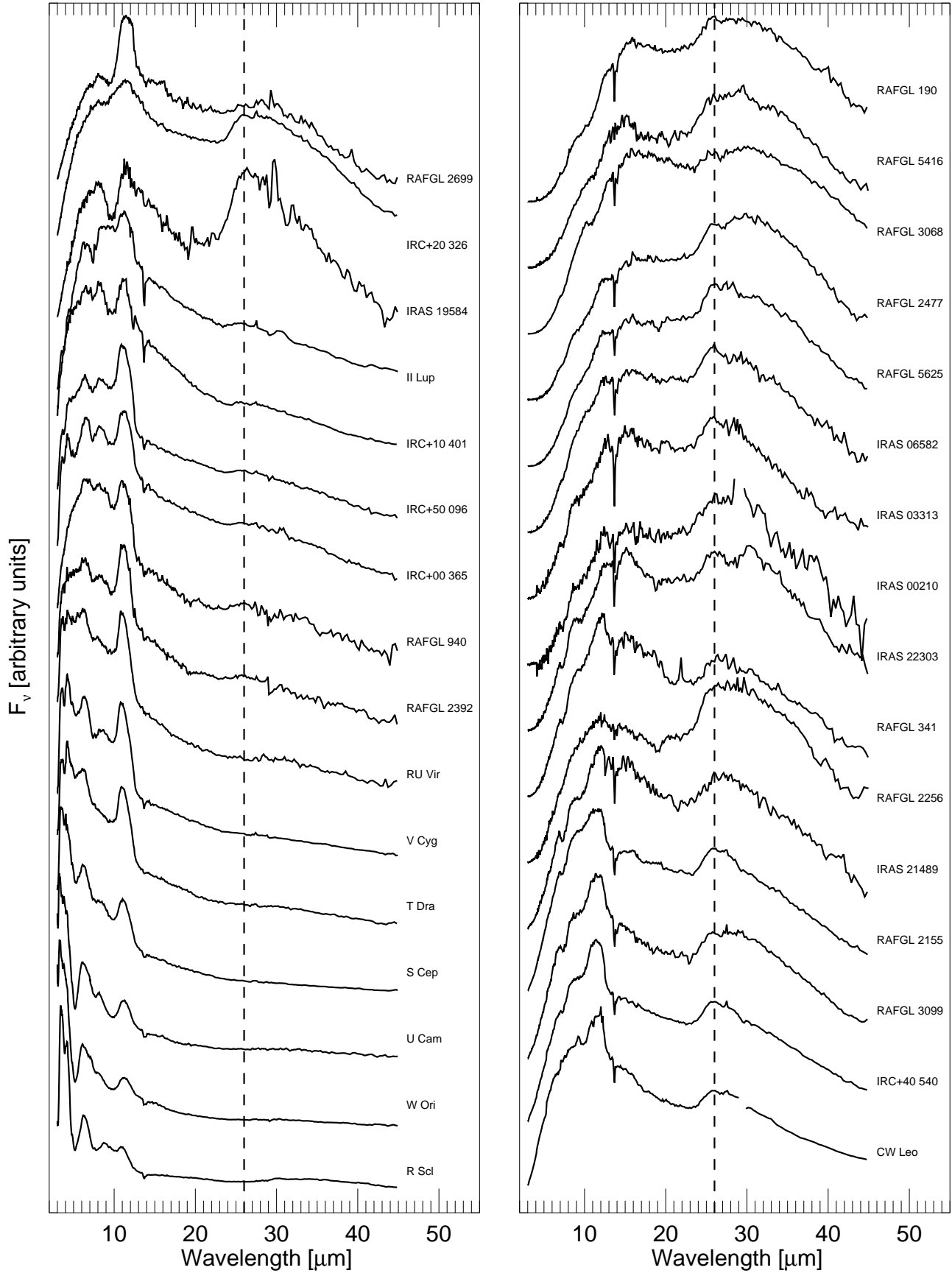}
  \caption{Overview of the spectra of carbon stars exhibiting the
    ``30'' $\mu$m feature. The spectra are ordered according to
    continuum temperature from high to low temperature, bottom to top,
    left to right. The dashed line marks $\lambda$=26 $\mu$m.}
  \label{fig:spectra}
\end{figure*}
\begin{figure*}[!hp]
  \includegraphics[width=17cm]{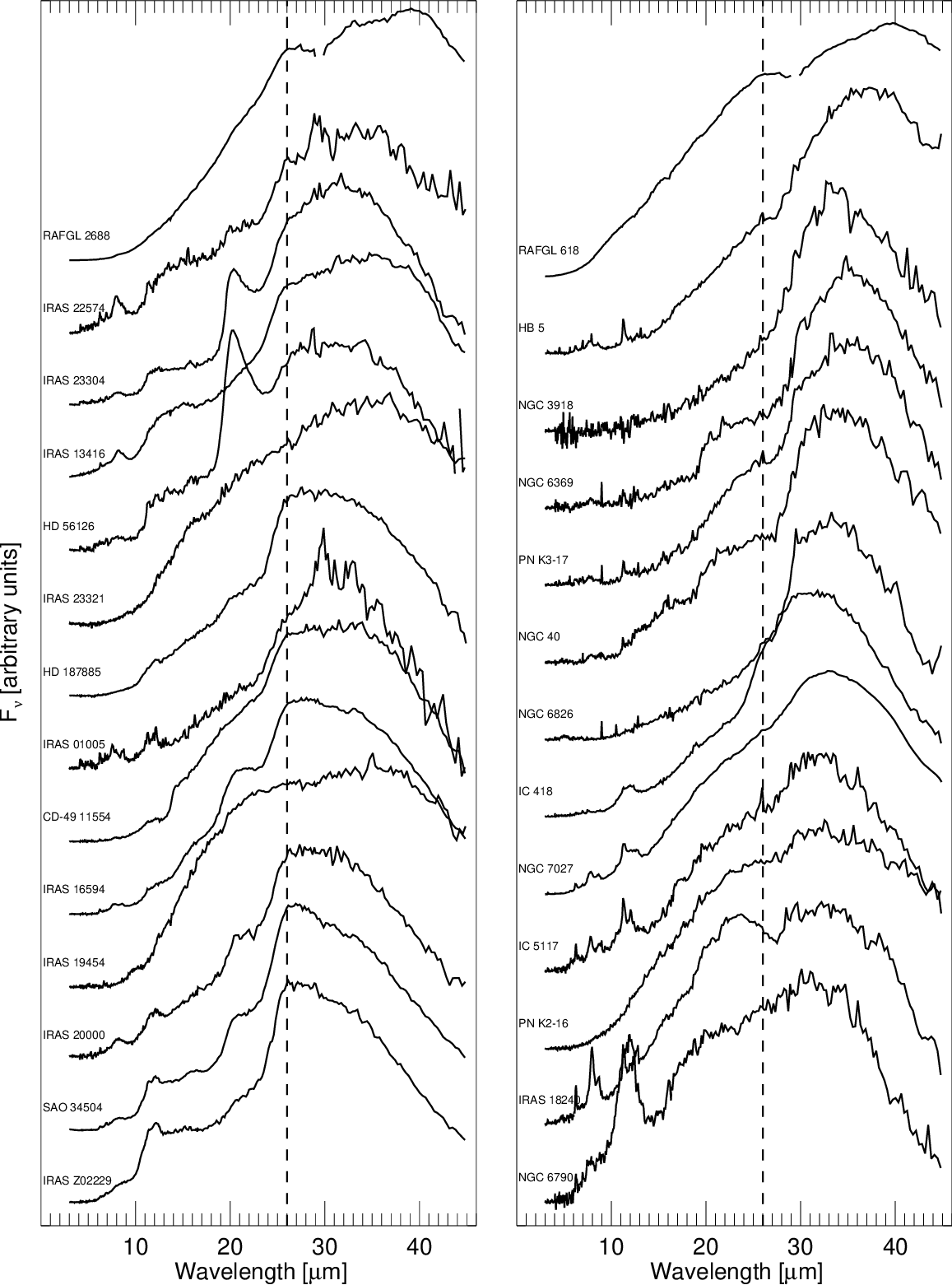}
  \caption{Overview of the spectra of post-AGB objects (left panel) and
    PNe (right panel) exhibiting the ``30'' $\mu$m feature. The
    spectra are ordered according to continuum temperature from high
    to low temperature, bottom to top. The dashed line marks
    $\lambda$=26 $\mu$m. The spectrum of RAFGL~618 although warmer
    than NGC~3918 is shown at the top of the PNe for clarity.}
  \label{fig:spectra2}
\end{figure*}
The resultant spectra for the sources that exhibit a ``30'' $\mu$m
feature are shown in Fig.~\ref{fig:spectra}. The SWS spectra of this
large group of objects show a spectacular range in colour temperature,
molecular absorption bands and solid state features.  The C-stars have
molecular absorption bands of C$_2$H$_2$ at 3.05, 7$-$8 and 14 $\mu$m,
of HCN at 7 and 14 $\mu$m, CO at 4.7 $\mu$m and C$_3$ at 4.8$-$6
$\mu$m. The sharp absorption band at 14 $\mu$m is due to C$_2$H$_2$
and HCN. There is an emission feature due to solid SiC at 11.4 $\mu$m.
In the reddest C-stars, we find the SiC in absorption.  We also find
evidence for a weak depression in the 14$-$22 $\mu$m range in the
reddest objects. This depression could be due to aliphatic chain
molecules like those found in RAFGL~618 \citep{2001ApJ...546L.127C}.

The post-AGBs and PNe exhibit many, sometimes broad solid state
emission features. In many sources we find emission due to polycyclic
aromatic hydrocarbons in the 3-15 $\mu$m range. There is a broad
plateau feature from 10$-$15 $\mu$m which may be due to hydrogenated
amorphous carbon \citep{1996ApJ...464..810G, 2001ApJ...554L..87K}.
Many post-AGBs and two PNe in the sample have a feature peaking at
20.1 $\mu$m, called the ``21'' $\mu$m feature in the literature.
Recently the carrier of this feature has been identified with TiC
\citep{2000Helden}. The feature at 23 $\mu$m found in IRAS~18240 and
PN K3-17 is likely due to FeS \citep{FeS_PNe_Hony}.  These absorption
and emission features have to be taken into account when determining
the profile of the ``30'' $\mu$m feature or the shape of the
underlying continuum.

Focusing on the ``30'' $\mu$m feature we can see variations in the
strength and shape of the band.  The most marked difference is however
a shift in the peak position going from 26 $\mu$m in some of the AGB
stars to 38 $\mu$m in the PNe.  The dashed line in
Fig.~\ref{fig:spectra} and \ref{fig:spectra2} indicates $\lambda$=26
$\mu$m. There are systematic changes in the appearance of the ``30''
$\mu$m feature from the C-stars to the PNe.  The feature in the
C-stars almost exclusively peaks at 26 $\mu$m.  There are some
exceptions like R~Scl. In the post-AGB sample, the feature is broader
and in some sources the feature peaks long ward of 26 $\mu$m. In the
PNe sample, there are \emph{no} sources that peak at 26 $\mu$m.
However, the appearance of a broad feature like the ``30'' $\mu$m
feature is sensitive to the shape of the underlying dust continuum,
especially since we have a sample with such a wide range of continuum
colour temperatures.

\section{Continuum}
\label{sec:continuum}
\begin{table*}
\caption{Measured properties. $T_\mathrm{cont}$, $p$ are 
    the parameters of the modified blackbody function fitted to the
    continuum. $\lambda_\mathrm{c,30}$ and P/C are the feature
    centroid position and peak over continuum ratio. $T_\mathrm{MgS}$
    is the derived temperature of the MgS grains.}
\begin{tabular}{l@{\ } l@{\ } l@{\ \ \ \ } l@{\ } l@{\ } l@{\ } l@{\ } l@{\,} | l@{\ } l@{\ } l@{\ \ \ \ } l@{\ } l@{\ } l@{\ } l@{\ } l@{\,}}
\hline
\hline
    %%Column header
     &
\multicolumn{2}{c}{cont.}&
\multicolumn{4}{c}{``30'' $\mu$m feature} &
     & &
\multicolumn{2}{c}{cont.}&
\multicolumn{4}{c}{``30'' $\mu$m feature}&
\\
    %%Column names
    Object&
    $T_\mathrm{cont}$&
    $p$&
    $\lambda_\mathrm{c,30}$&
    fwhm&
    flux&
    P/C&
    $T_\mathrm{MgS}$ &
    Object&
    $T_\mathrm{cont}$&
    $p$&
    $\lambda_\mathrm{c,30}$&
    fwhm&
    flux&
    P/C&
    $T_\mathrm{MgS}$
\\
    %% Column units
    &
    [K]&
    &
    [$\mu$m]&
    [$\mu$m]&
    [W/m$^2$]&
    &
    [K]&
    &
    [K]&
    &
    [$\mu$m]&
    [$\mu$m]&
    [W/m$^2$]&
    &
    [K]
\\
\hline
NGC 40          &  150 &  0     &  33.6 & 10.1 & 5.9e-13 &  0.7 & 110    &T Dra           & 1210 &  0     &  30.2 & 10.1 & 4.8e-13 &  0.4 & 200\\
IRAS 00210      &  285 &  0.5   &  28.4 & 10.7 & 6.4e-13 &  0.8 & 300    &RAFGL 2155      &  460 &  0     &  28.8 &  8.2 & 5.7e-12 &  0.6 & 400\\
IRAS 01005      &  130 &  1     &  30.0 & 11.1 & 6.6e-13 &  1.5 & 220    &IRAS 18240      &  160 &  1     &  32.8 & 13.1 & 1.0e-12 &  1.0 & 130\\
HV Cas          & 1040 &  0.2   &  33.5 & 10.6 & 1.5e-13 &  0.3 & 100:   &IRC+00 365      &  910 & -0.3   &  28.6 & 11.7 & 1.9e-12 &  0.4 & 500\\
RAFGL 190       &  275 &  0     &  30.9 & 13.0 & 1.6e-12 &  0.3 & 180    &RAFGL 2256      &  390 &  0     &  29.5 & 12.0 & 1.9e-12 &  1.0 & 350\\
R Scl           & 2605 & -0.2   &  33.2 & 13.9 & 1.1e-12 &  1.1 & 90     &K3-17           &  100 &  1     &  34.1 & 11.5 & 1.0e-12 &  0.9 & 90\\
IRAS Z02229     &  235 &  0     &  29.1 & 10.1 & 8.3e-12 &  1.7 & 300    &IRC+10 401      &  765 &  0     &  30.0 & 10.0 & 2.0e-12 &  0.3 & 300\\
RAFGL 341       &  380 &  0     &  29.8 &  9.4 & 9.4e-13 &  0.4 & 250    &IRAS 19068      & 1165 & -0.7   &  28.5 & 10.1 & 2.0e-13 &  0.4 & 500:\\
IRC+50 096      &  855 & -0.2   &  28.8 &  9.2 & 1.9e-12 &  0.3 & 500    &NGC 6790        &  290 &  0     &  29.8 & 15.6 & 9.8e-13 &  1.4 & 300\\
IRAS 03313      &  325 &  0     &  28.6 &  7.8 & 5.4e-13 &  0.4 & 300    &RAFGL 2392      &  890 &  0     &  27.7 &  8.6 & 3.4e-13 &  0.5 & 500\\
U Cam           & 1775 &  0     &  31.9 & 11.8 & 3.9e-13 &  0.6 & 150    &NGC 6826        &  150 &  0     &  32.7 & 10.5 & 1.1e-12 &  2.0 & 120\\
RAFGL 618       &  235 & -1     &  38.0 & 10.9 & 5.4e-12 &  0.2 & 40$^a$ &IRAS 19454      &  140 &  1     &  36.3 & 13.1 & 6.4e-13 &  0.3 & 50\\
W Ori           & 2450 &  0     &  31.3 &  8.4 & 3.1e-13 &  0.4 & 150    &HD 187885       &  175 &  0     &  29.6 & 10.8 & 5.2e-12 &  1.0 & 200\\
IC 418          &  120 &  1     &  30.8 & 11.3 & 5.5e-12 &  0.9 & 180    &RAFGL 2477      &  290 &  0     &  30.7 & 12.5 & 2.3e-12 &  0.6 & 170\\
V636 Mon        & 1215 &  0     &  29.8 & 10.1 & 1.7e-13 &  0.2 & 250:   &IRAS 19584      &  580 &  0     &  28.1 &  7.5 & 8.5e-13 &  1.5 & 400\\
RAFGL 940       &  810 &  0     &  28.2 & 10.2 & 3.5e-13 &  0.5 & 500    &IRAS 20000      &  210 &  0     &  29.4 & 12.1 & 2.5e-12 &  1.5 & 300\\
IRAS 06582      &  315 &  0     &  29.5 & 10.3 & 1.1e-12 &  0.4 & 300    &V Cyg           & 1110 &  0     &  30.5 & 11.5 & 1.3e-12 &  0.3 & 200\\
HD 56126        &  170 &  0     &  30.0 & 12.0 & 2.9e-12 &  0.8 & 150    &NGC 7027        &  125 &  1     &  32.8 & 11.0 & 1.7e-11 &  0.4 & 110\\
CW Leo          &  535 &  0     &  28.6 &  8.8 & 2.7e-10 &  0.6 & 400    &S Cep           & 1340 &  0.1   &  31.2 &  9.4 & 4.4e-13 &  0.2 & 130\\
NGC 3918        &   90 &  1     &  33.3 &  8.5 & 7.1e-13 &  1.0 & 120    &RAFGL 2688      &  200 & -1     &  31.1 & 10.4 & 5.9e-11 &  0.3 & 70$^a$\\
RU Vir          & 1045 &  0     &  30.4 & 10.1 & 5.3e-13 &  0.6 & 180    &RAFGL 2699      &  540 &  0     &  29.0 & 11.4 & 5.9e-13 &  0.7 & 300\\
IRAS 13416      &  115 &  1     &  31.6 & 15.8 & 2.8e-12 &  0.4 & 200$^a$&IC 5117         &  130 &  1     &  31.2 &  9.7 & 7.3e-13 &  0.6 & 150\\
II Lup          &  625 &  0     &  29.5 & 10.1 & 3.9e-12 &  0.3 & 400    &RAFGL 5625      &  300 &  0     &  30.3 & 11.8 & 4.4e-12 &  0.4 & 200\\
V Crb           & 1430 &  0     &  30.4 & 10.1 & 1.8e-13 &  0.3 & 150:   &IRAS 21489      &  415 &  0     &  29.3 &  9.7 & 1.1e-12 &  0.6 & 350\\
K2-16           &  155 &  0.5   &  34.4 & 12.0 & 3.4e-13 &  0.3 & 80     &SAO 34504       &  210 &  0     &  29.1 & 10.3 & 1.3e-11 &  2.0 & 250\\
IRAS 16594      &  140 &  1     &  29.8 & 12.1 & 9.9e-12 &  0.9 & 250    &IRAS 22303      &  345 &  0     &  30.3 & 10.5 & 1.0e-12 &  0.7 & 300\\
NGC 6369        &  100 &  1     &  34.6 & 10.1 & 9.5e-13 &  1.1 & 90     &IRAS 22574      &  160 &  0     &  31.2 & 13.6 & 5.9e-13 &  0.4 & 150\\
IRC+20 326      &  770 & -0.7   &  29.1 & 10.2 & 7.4e-12 &  0.5 & 300    &RAFGL 3068      &  290 &  0     &  32.4 & 14.7 & 8.4e-12 &  0.4 & 120\\
CD-49 11554     &  140 &  1     &  30.2 & 14.0 & 4.7e-12 &  0.7 & 200$^a$&RAFGL 3099      &  470 &  0     &  29.5 & 10.9 & 2.6e-12 &  0.7 & 400\\
HB 5            &  120 &  0     &  35.5 & 11.5 & 1.0e-12 &  0.4 & 70     &IRAS 23304      &  115 &  1     &  30.1 & 13.4 & 2.3e-12 &  1.1 & 250\\
RAFGL 5416      &  290 &  0     &  30.4 & 12.5 & 2.2e-12 &  0.5 & 220    &IRAS 23321      &  175 &  0     &  34.5 & 13.3 & 6.6e-13 &  0.3 &  70\\
IRC+40 540      &  485 &  0     &  28.6 &  9.1 & 8.9e-12 &  0.6 & 400    &                &      &        &       &      &\\
\hline
\multicolumn{16}{c}{non detections}\\
\hline
R For           & 1215 &  0     &  -    &   -  &$<$1e-14 &$<$0.1&  -     &T Lyr           & 3305 &  0     &  -    &   -  &$<$1e-14 &$<$0.1&  -\\
SS Vir          & 2040 &  0     &  -    &   -  &$<$1e-14 &$<$0.1&  -     &S Sct           & 2105 &  0     &  -    &   -  &$<$4e-14 &$<$0.3&  -\\
Y CVn           & 2200 &  0     &  -    &   -  &$<$2e-13 &$<$0.2&  -     &V Aql           & 3665 & -0.3   &  -    &   -  &$<$1e-14 &$<$0.1&  -\\
RY Dra          & 2525 &  0     &  -    &   -  &$<$1e-13 &$<$0.2&  -     &V460 Cyg        & 2875 &  0     &  -    &   -  &$<$5e-14 &$<$0.5&  -\\
C* 2178         & 1110 &  0     &  -    &   -  &$<$1e-13 &$<$0.5&  -     &PQ Cep          & 1625 &  0     &  -    &   -  &$<$1e-14 &$<$0.1&  -\\
V1079 Sco       & 3085 & -0.5   &  -    &   -  &$<$5e-14 &$<$0.2&  -     &TX Psc          & 3105 &  0     &  -    &   -  &$<$3e-14 &$<$0.1&  -\\
\hline
\hline
\end{tabular}
$^a$Temperature determination uncertain due to optically thick MgS emission.
\label{tab:properties}
\end{table*}
\begin{figure}
  \includegraphics[width=8.8cm]{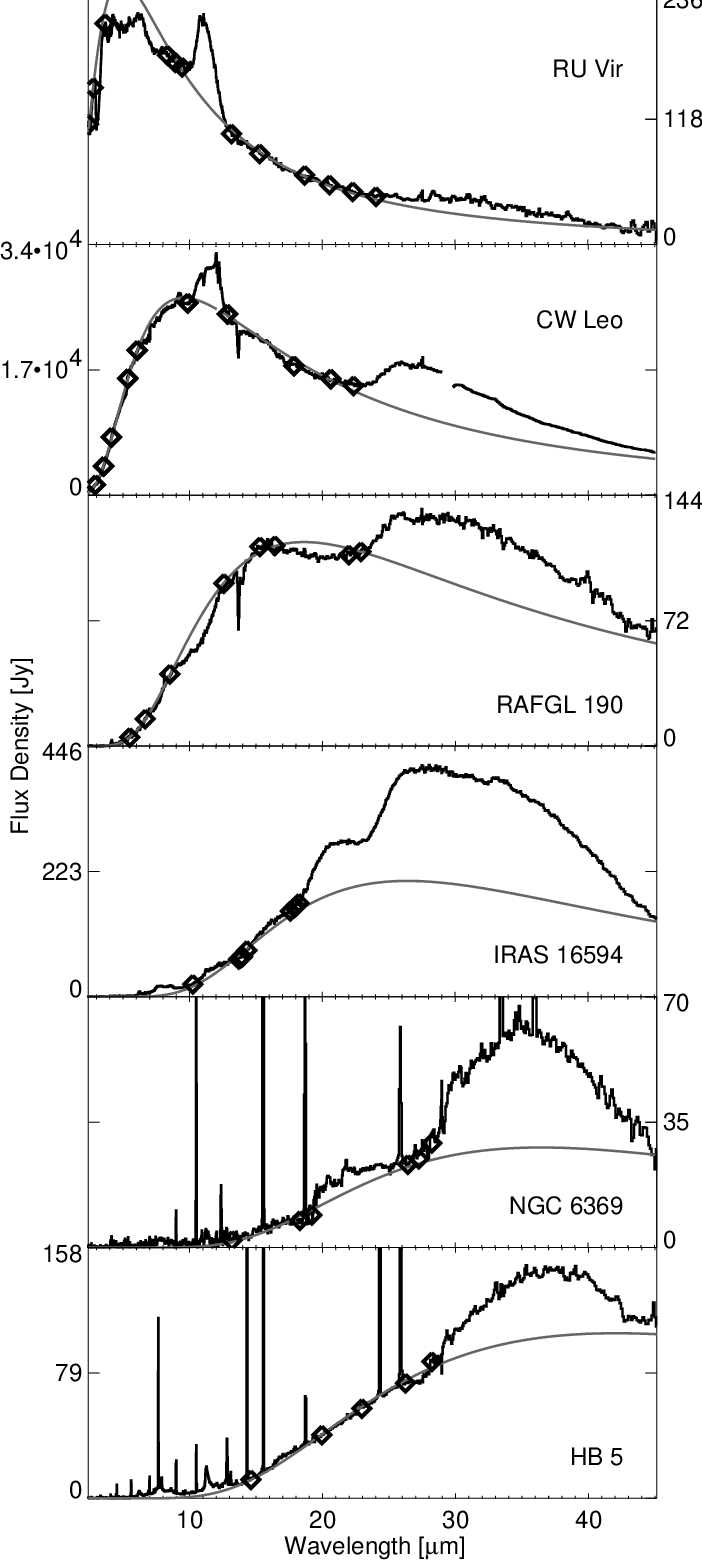}
  \caption{Examples of the fitted continuum. We show the spectra
    (black line), the selected continuum points (diamonds) and the
    fitted modified blackbody (grey line).}
  \label{fig:contfit}
\end{figure}
In order to extract the profiles of the ``30'' $\mu$m features and
compare them from source to source, we model the underlying continuum
due to the emission of other circumstellar (CS) dust components.
First, we present the way we construct these continua and in
Sect.~\ref{sec:profiles}, we discuss the resulting profiles.

To model the underlying continuum we use a simplified approach. We
represent the continuum with a single temperature modified blackbody,
\begin{equation}
F(\lambda)=A \times B(\lambda,T) \times {\lambda}^{-p},
\label{eqn:modifiedblackbody}
\end{equation}
where $\lambda$ is the wavelength, $F(\lambda)$ is the flux density of
the continuum, $B(\lambda,T)$ is the Planck function of temperature
$T$, $p$ is the dust emissivity index and $A$ is a scaling factor.

We have chosen this approach to estimate the continuum over doing a
radiative transfer calculation for reasons of simplicity. The bulk of
the CS dust around these sources consists of some form of amorphous
carbon grains that do not exhibit sharp emission features in the
wavelength range of interest. Therefore, a radiative transfer
calculation will not yield extra insight into the shape or strength of
the continuum while introducing many more modelling parameters. This
method has the advantage that we can compare the feature in such a
diverse group of sources in a consistent way. Of course
Eq.~(\ref{eqn:modifiedblackbody}) does not directly allow us to
incorporate important effects such as optical depth or temperature
gradients. However varying the $p$-parameter can mimic these effects
to some extent.

The $p$-parameter reflects the efficiency with which the dust grains
can emit at wavelengths larger than the grain size. Reasonable values
of $p$ in the region of interest are between 1 and 2.  Crystalline
materials have this value close to 2 and amorphous materials have a
$p$-value between 1 and 2, while layered materials have an emissivity
index close to 1. A temperature gradient in the dust shell will result
in a broader spectral energy distribution (SED). This is mimicked by a
lower value of $p$. Likewise an optically thick dust shell will result
in a broader SED, which again can be reproduced by reducing the value
of $p$.

We use a ${\chi}^{2}$ fitting procedure to determine the values of $T$
and $p$ fitted to selected continuum points in the ranges 2$-$22
$\mu$m.  If available we also use the LWS spectra to verify the
continuum at the long wavelength end of the ``30'' $\mu$m feature.
The 50$-$100 $\mu$m continuum gives an even stronger
  constraint on the value of $p$.  For most cases the resultant
continuum runs through the 45 $\mu$m region of the SWS spectrum. A
remarkable exception to this is the spectrum of RAFGL~3068. The 2$-$24
$\mu$m spectrum is well fitted with a single 290 K Planck function.
However we find a large excess of this continuum at 45 $\mu$m and the
available LWS spectrum is not well represented in level or slope.
Possibly this is due to the optically thick dust shell or a biaxial
dust/temperature distribution.

The values for $T$ and $p$ are listed in Table~\ref{tab:properties}.
One remarkable fact is that the C-stars are well fitted by a single
temperature Planck function over the complete wavelength range of SWS.
The IR SEDs of the post-AGBs and PNe are in general less broad and
many sources are better fitted with a $p$-value of 1. We stress
however that the derived $p$ values cannot be used to constrain the
crystal structure or the average size of the dust grains in view of
the aforementioned effects of temperature gradients and optical depth.

\begin{figure}
  \includegraphics[width=8.8cm]{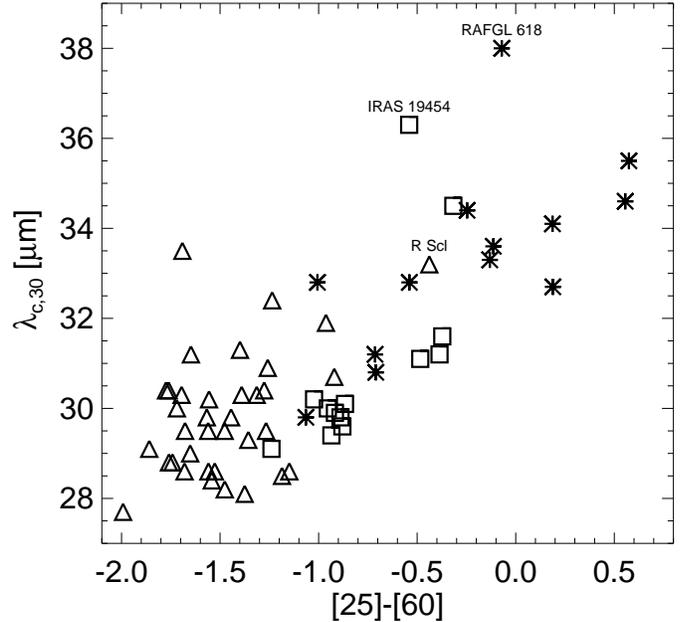}
  \caption{Centroid of the ``30'' $\mu$m feature  with respect to the
    [25]-[60] colour. The symbols are like in
    Fig.~\ref{fig:irascolours}. There is a clear trend for the
    centroid position of the ``30'' $\mu$m feature to move to longer
    wavelengths the redder the object is.}
  \label{fig:peakshift}
\end{figure}
\section{Profiles}
\label{sec:profiles}
\begin{figure}
  \includegraphics[width=8.8cm]{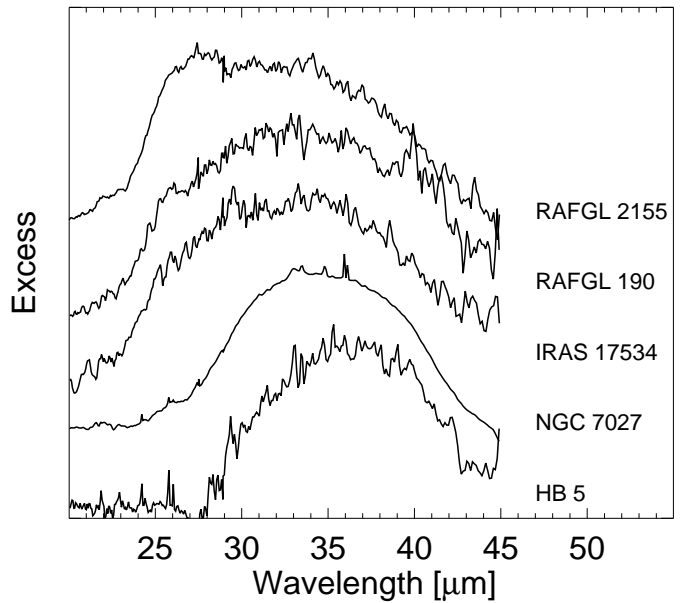}
  \caption{Emissivities for different sources as deduced from the
    ISO spectra. There are large differences in the profile of the
    excess emission. Notice the shift in peak position and change in
    width of the circumstellar ``30'' $\mu$m feature going from the
    C-stars to post-AGBs to PNe (top to bottom).}
  \label{fig:emissivity}
\end{figure}
Using the continua defined in this way, we measure the following
properties of the ``30'' $\mu$m feature: the centroid wavelength
($\lambda_\mathrm{c,30}$), i.e., the wavelength where the integrated
flux in the feature at either side is equal; the full width at half
maximum (fwhm); the flux in the ``30'' $\mu$m feature and the height
of the peak of the ``30'' $\mu$m feature \emph{after continuum}
subtraction over the continuum ratio, i.e., the peak to continuum
ratio (P/C). The measured values are listed in
Table~\ref{tab:properties}. We list upper limits for P/C and the flux
for the sources without detection.  In Fig.~\ref{fig:peakshift}, we
show the relation between the [25]-[60] colour and
$\lambda_\mathrm{c,30}$. There is a clear reasonably smooth trend for
the feature to move to longer wavelengths with redder IRAS colours.
This indicates that the temperature of the dust is an important
parameter in determining the profile of the ``30'' $\mu$m feature
since [25]-[60] is a direct measure of the dust temperature provided
that the dust composition in the different sources is similar.

We first try to remove the effect of temperature by dividing by the
continuum; a method that is commonly applied. Using the modelled
continua as described in Sect.~\ref{sec:continuum}, we convert the
observed features to relative excess emission by dividing by the
continuum and subtracting 1.

If the feature emission is optically thin and the temperature of the
carriers of the feature is equal to the continuum temperature the
derived excess emissions are proportional to the absorptivity
($\kappa_\mathrm{abs}$) of the carrier and if the carrier is the same
in these sources then the derived band shape should be the same for
all sources.  However, we find large variations in the derived
profiles.  In Fig.~\ref{fig:emissivity}, we show some examples of the
derived profiles. Most notable are variations in peak position and the
appearance around 26 $\mu$m. Such changes, albeit within a smaller
range of feature peak positions have led other authors
\citep{2000ApJ...530..408V,Volk2002} to conclude that the ``30''
$\mu$m feature is composed of two features and the observed variations
are due to varying relative contributions of these two components. One
key question is: ``What possible causes could there be for the
observed large variations in band shape?''. We discuss three
possibilities below. First, optical depth effects. Second, temperature
effects. Finally, we discuss multiple band carriers.

The optically thin assumption most likely holds because the optical
depth in the circumstellar shell strongly decreases towards longer
wavelengths. Note, in this respect that the ``30'' $\mu$m feature is
never found in absorption (however, see also
Sect.~\ref{sec:opticallythick}). Hence, optical depth effects are not
responsible for the observed profile variations.

Whether the temperature of the amorphous carbon grains (defining the
shape of the continuum) and the temperature of the ``30'' $\mu$m
carrier are equal is very uncertain.  The temperature of a dust grain
in a circumstellar envelope is determined by the distance to the star,
the absorption properties in the wavelength range where the star or
the dust shell emits light and the grain size. In case the temperature
of the grains species responsible for the continuum and the ``30''
$\mu$m emission feature are not the same, the resulting excess
profiles will also not be the same from source to source \emph{even if
  the carrier of the band is the same}. The differences will be very
pronounced when the emission feature is broad. In this case systematic
difference between sources are bound to occur in league with the
strongly changing continuum temperature.  Thus, the temperature of the
carrier of the ``30'' $\mu$m feature is an important parameter that
determines the profile of the emission.

There may be multiple carriers involved as discussed before.  In this
case the feature near 26 $\mu$m dominates in the warmest objects while
the cooler objects are more and more dominated by emission towards 35
$\mu$m. However, this scenario has its difficulties since it would
require changes in the composition of the dust in the relatively
dispersed and cold nebular surroundings of a post-AGB object or even
during the PN phase. Such chemical changes can only occur extremely
slowly, if at all.

Lastly, variations in grain shape or variations in shape distribution
can influence the emission profiles. The optical properties of
materials with a high value of the refractive index are sensitive to
the grain shape. Variations in the shape distribution will lead to
variations in the profiles.

In our analysis, we will focus on explaining the profile variations
with temperature variations and the effects of variations in the shape
distribution of the emitting dust grains.
\begin{figure}[!t]
  \includegraphics[width=8.8cm]{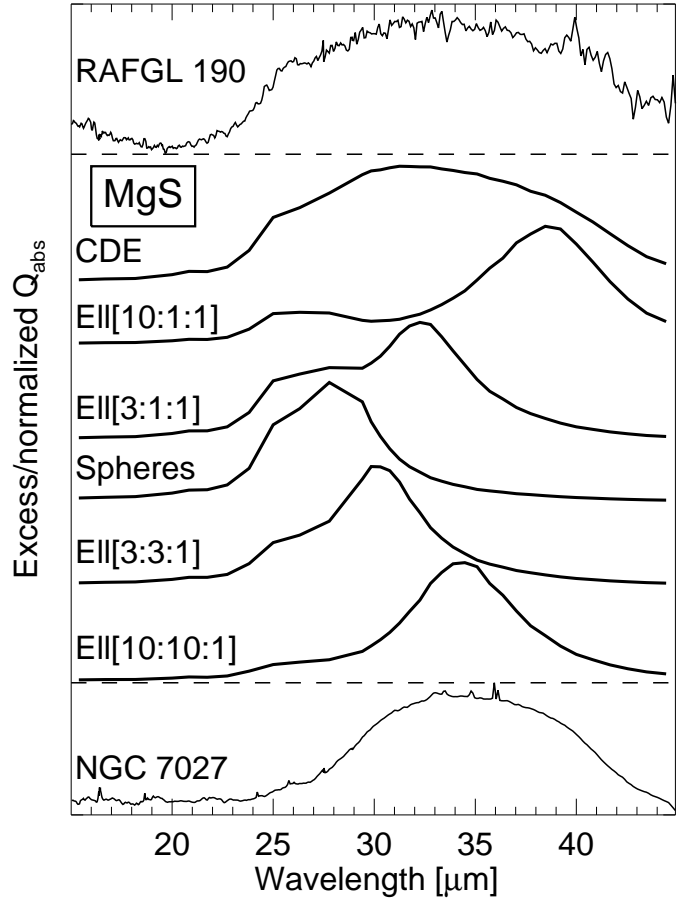}
  \caption{The effect of grain shape. We show the absorptivity of MgS
    as a function of grain shape. The numbers between brackets refer
    to the axis ratios of the elliptical grains. For comparison we
    show the derived emissivities of RAFGL~190 and NGC~7027.}
  \label{fig:featurespectra}
\end{figure}

\section{MgS}
\label{sec:mgs}
\begin{figure}[!t]
  \includegraphics[width=8.8cm]{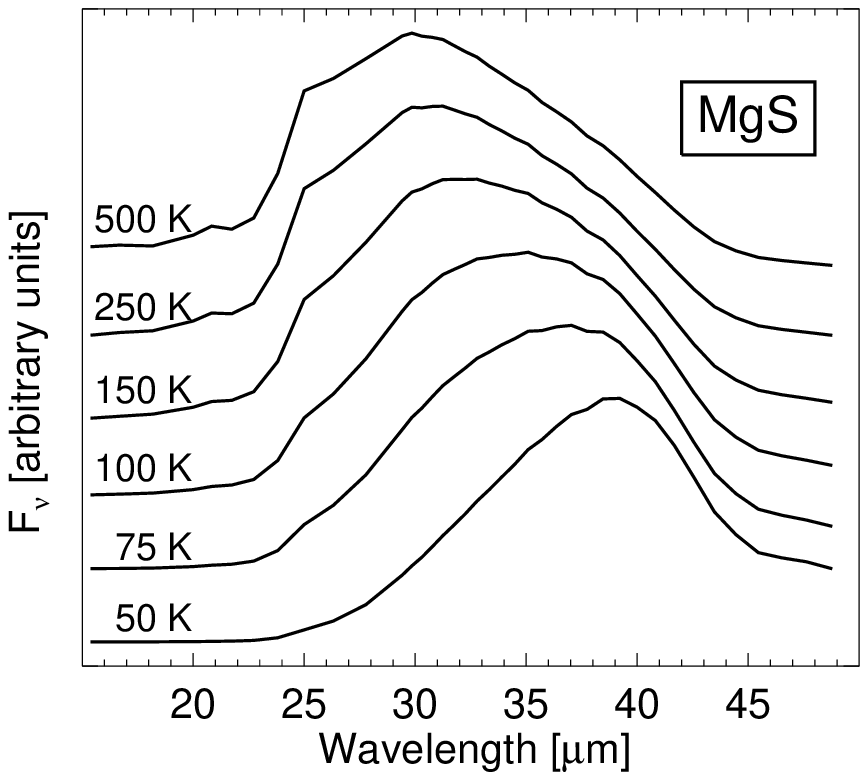}
  \caption{The effect of grain temperature on the MgS emission
    feature. We fold the $\kappa_\mathrm{abs}$ of MgS in a 
      continuous distribution of ellipsoids (CDE)} shape distribution
    with a Planck function of different temperatures.  The shape and
    position of the feature are modified substantially.
  \label{fig:mgstemp}
\end{figure}
Since, it has been demonstrated for a few sources that MgS is a viable
candidate \citep{1994ApJ...423L..71B, 1999A&A...344..918J,
  1999A&A...345L..39S}, we first test MgS as a possible candidate for
the carrier of the feature. With the large sample of good quality
spectra in this study we are able to test this possible identification
systematically in a large population of evolved objects.

As explained above we cannot derive \emph{a priori} information on the
temperature of the ``30'' $\mu$m carrier from the observations. Our
knowledge is further limited by the fact that even for some of the
candidate materials like MgS or FeS the optical properties are
measured only in a limited wavelength range. We lack measurements in
the UV, optical and near-IR range, which may well dominate the dust
heating. We have decided to test the MgS identification, leaving the
grain temperature as a free parameter. We adopt the method we describe
below.

\subsection{Material}
We use the optical constants as published by
\citet{1994ApJ...423L..71B}. Of the materials they measured,
Mg$_{0.9}$Fe$_{0.1}$S is closest to pure MgS. The real and imaginary
part of the refractive index ($n$ and $k$ values) are given from
10-500 $\mu$m.

\subsection{Shapes}
From the $n$ and $k$ values we can calculate the absorption
cross-sections for various grain shapes and shape distributions in the
Rayleigh limit following \citet[][chap.~5, 9 and 12]{BohrenHuffman}.
The absorption cross-section of MgS around 30 $\mu$m is very sensitive
to the grain shape. In Fig.~\ref{fig:featurespectra}, we show the
results of using different grain shapes on the absorption properties
of MgS.  We use a continuous distribution of ellipsoids (CDE) for the
grain shapes.  This shape distribution was used by
\citet{1994ApJ...423L..71B} and gave a good fit to the ``30'' $\mu$m
feature observed in CW~Leo. The same shape distribution was further
used by \citet{1999A&A...344..918J} and \citet{1999A&A...345L..39S}.
They found reasonable fits for the two sources they study.  As can be
clearly seen in Fig.~\ref{fig:featurespectra} when comparing the
spheres with the CDE calculations, the feature broadens and the peak
position shifts to longer wavelengths using the CDE shape
distribution. The width of the feature calculated using CDE matches
that of the observed ``30'' $\mu$m feature well (e.g.  RAFGL~190 in
Fig.~\ref{fig:featurespectra}, see also Fig.~\ref{fig:fits}).

\subsection{Temperature}
To estimate the MgS temperature ($T_\mathrm{MgS}$) we use the
continuum \emph{subtracted} spectra with the continuum as derived in
Sect.~\ref{sec:continuum}. The emission from MgS grains is calculated
using the $\kappa_\mathrm{abs}$ folded with a Planck function with the
temperature of the grain. Due to the smooth and broad shape of the
resonance the profile of the emission is very sensitive to
$T_\mathrm{MgS}$. In particular the peak position changes strongly
with $T_\mathrm{MgS}$.  This allows us to estimate $T_\mathrm{MgS}$
from the continuum subtracted profiles. This method is most sensitive
for $T_\mathrm{MgS}$ $<$ 300~K. Above 300~K, further changes in the
profile are more subtle since the major part of the feature falls in
the Rayleigh-Jeans domain of the Planck function.

We use this temperature estimate and the observed band strength in the
continuum subtracted spectra to synthesise a MgS feature in order to
compare with the astronomical spectra. In conclusion, we adopt MgS
with a CDE shape distribution and allow both the strength and the
temperature of the MgS grains to vary with respect to the underlying
continuum.

\section{Model results}
\label{sec:modelresult}
\begin{figure}
  \includegraphics[width=8.8cm]{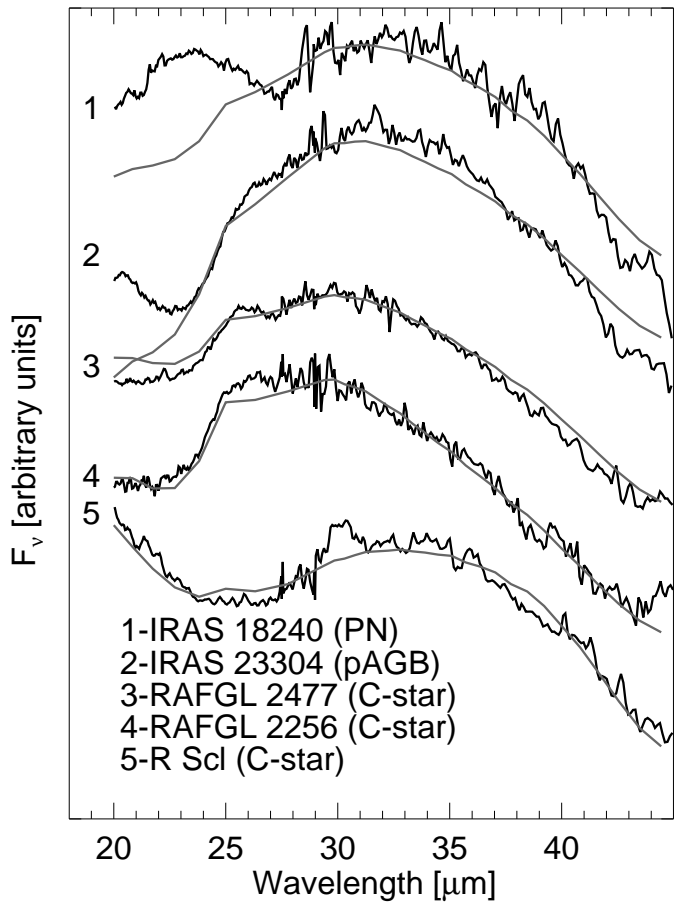}
  \caption{Examples of spectra that are very well fitted with MgS in a
    single temperature, CDE shape distribution. The black lines
    represent the data and the grey line the model. The different
    sources have been offset for clarity. The excess around 23 $\mu$m
    in IRAS~18240 is due to FeS \citep{keller_nature,FeS_PNe_Hony}.
    Notice how the model is able to explain the profile of the ``30''
    $\mu$m feature found in the full range of objects in our sample.}
  \label{fig:perfectfit}
\end{figure}
We show in Fig.~\ref{fig:fits} the observed spectra, the composite of
the continuum and the synthetic MgS feature and the residuals after
subtracting the MgS feature. The fits are very satisfactory in 50 out
of 63 cases. In $\sim$25 sources the synthetic spectra obtained with
this very simple model are able to explain the detailed profile of the
``30'' $\mu$m feature very well. The onset and range of the feature
and even the slight depression between 26$-$30 $\mu$m are reproduced
by the model. We show a zoomed view of the 30 $\mu$m region of a few
sources that are very well fitted by this simple model in
Fig.~\ref{fig:perfectfit}.  Notice the different apparent shapes of
the feature that the model is able to explain. \citet{Volk2002}
discuss the ``30'' $\mu$m feature in IRAS~23304 and find that they
need 2 separate unidentified components in order to understand the
shape of the feature.  Fig.~\ref{fig:perfectfit} illustrates that this
is not necessarily required.

Examining the complete sample of observed ``30'' $\mu$m features and
the synthetic spectra, we find there are some systematic deviations.
In the sample of C-stars and post-AGB objects there are numerous
examples where the major part of the ``30'' $\mu$m feature is
explained well by our model, but the observed spectra show excess
emission in the 26 $\mu$m region. The excess is not accounted for using
our CDE fits. The most extreme case is IRAS~19584 but several sources
exhibit the same behaviour. In Sect.~\ref{sec:26-mum-excess}, we
discuss the origin of this discrepancy.

In some cases the synthetic spectra over-predict the flux at the
longest wavelengths. This can be due to the very simplistic method we
have used to estimate the continuum level. As the dust optical depth
decreases with increasing wavelengths, the continuum level estimated
from shorter wavelengths might over-predict the true continuum level.
Note, however, that the discrepancies between the modelled and the
observed spectra in the 26 $\mu$m region and the 40 $\mu$m region
cannot be considered completely independently. If we were to weaken
the strength of the MgS feature this would yield a better fit around
45 $\mu$m but would increase the discrepancy around 26 $\mu$m.  We
also note that MgS produces a weak continuum contribution at 45 $\mu$m
(see for example Fig.~\ref{fig:mgstemp}). This continuum contribution
is already taken into account when fitting the overall continuum, but
it is still present in the calculated MgS contribution. Therefore, our
model may slightly over-predict the fluxes near 45 $\mu$m.

As a class, the spectra of most PNe show another systematic
difference.  The peak position of the ``30'' $\mu$m feature lies in
general at longer wavelengths than in the post-AGB sources. This is in
accordance with the picture of a slowly expanding and cooling dusty
envelope. We can simulate the same shift in peak position using MgS
grains. However the fits we obtain fail to reproduce the relatively
narrow width of the observed profile. We discuss this deviation of the
profiles in Sect.~\ref{sec:pne-profiles}.

There are 4 sources in the sample that have a broader ``30'' $\mu$m
feature than can be fit by our simple model. Of these sources,
IRAS~13416 and CD-49~11554 show a slightly flattened and broadened
feature while in the cases of RAFGL~618 and RAFGL~2688 the feature is
very broad with a depression around 30 $\mu$m. The latter sources are
known to have a very large dust column along the line of sight.  Most
likely the feature shape is due to optical depth effects. We discuss
these sources further in Sect.~\ref{sec:opticallythick}.

Despite these systematic deviations it is clear that our simple model
is able to explain the profile of the ``30'' $\mu$m feature in good
detail in a very wide range of objects. We conclude that the carrier
of the ``30'' $\mu$m feature in the C-stars and post-AGB objects is
solidly identified with MgS and that \emph{the variations in peak
  position reflect differences in grain temperature}.

\subsection{26 $\mu$m excess}
\label{sec:26-mum-excess}
\begin{figure}
  \includegraphics[width=8.8cm]{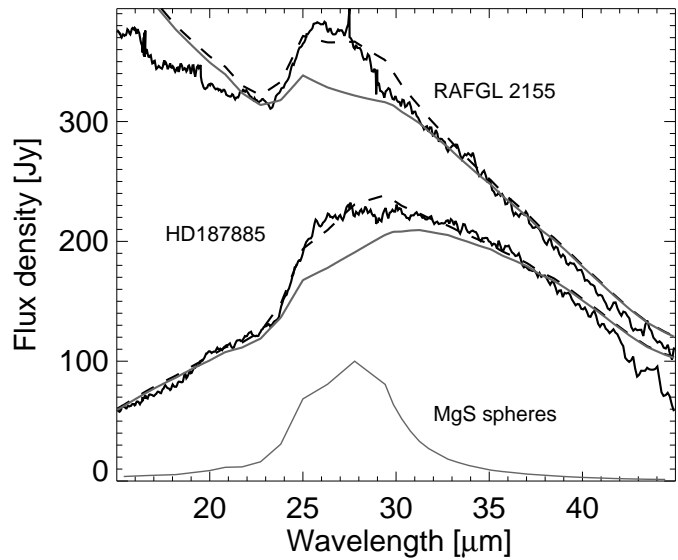}
  \caption{Some examples of sources with a 26 $\mu$m excess compared
    to the model spectra. We show the data (full black line) the model
    using the CDE calculation (full grey line) and the composite model
    using both spherical MgS grains and the CDE calculation (dashed
    line). We show below the emission of spherical MgS grains at a
    temperature of 250 K. The continuum in RAFGL~2155 runs above the
    observed spectrum at the shortest wavelengths, this may be due to
    molecular absorptions (see Sect.~\ref{sec:spectra})}
  \label{fig:26excess}
\end{figure}
As discussed above we find about 25 sources that are very well fitted
by our simple model. Using similar parameters we also find about 25
sources which show an excess near 26 $\mu$m. Evidently, there is a
contribution from an additional dust component in the latter sources.
In considering this additional dust component we find that MgS itself
is the best candidate. The wavelength region where this excess occurs
(26 $\mu$m) is also the wavelength where the generic main resonance of
MgS occurs. Spherical MgS grains exhibit a resonance at 26 $\mu$m. In
our model, the MgS profile is broader and peaks at a longer wavelength
because we use the calculated absorption cross-sections in a CDE shape
distribution. There is no a priori reason why the shape distribution
should be close to this distribution. In this respect, it is rather
surprising that our simple model works so well for so many sources. We
can simulate a different shape distribution by adding the
contributions of spherical grains to the CDE profile. This approach is
permitted since we assume optically thin emission and thus the
contribution of different components add linearly.  In
Fig.~\ref{fig:26excess}, we show the result of such a composite model
for one C-star and for one post-AGB object. The relative amounts of
spherical grains added to the model is $\sim$35 per cent. We keep the
temperature of the spherical and the CDE grains the same as in the
initial model. As can be seen the spherical grains contribute at the
position where our initial model fails. We conclude that variations in
the distribution over grain shapes can explain the variations we
observe in the profile of the ``30'' $\mu$m feature.

\subsection{Optically thick shells}
\label{sec:opticallythick}
\begin{figure}
  \includegraphics[width=8.8cm]{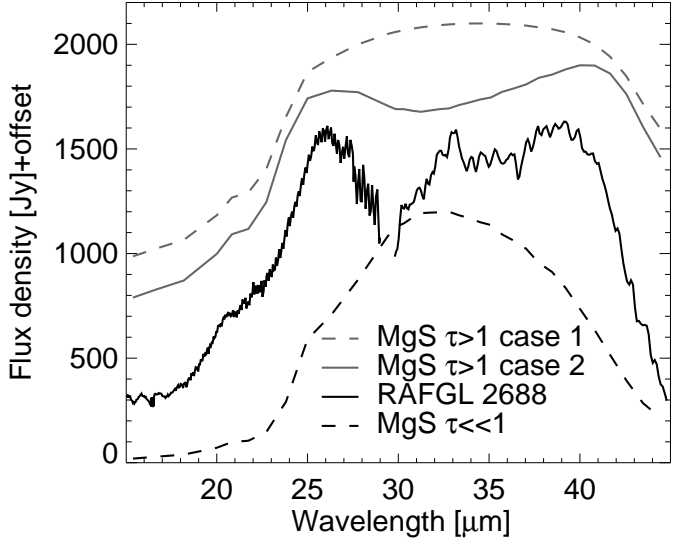}
  \caption{The effects of optical depth of the profile of the MgS
    emission. We show the absorbed MgS emission following
    Eq.~(\ref{eq:opticallythick1}) and (\ref{eq:opticallythick2})
    (grey dashed and solid lines), the profile of the ``30'' $\mu$m
    feature of RAFGL~2688 (solid black line) and the optically thin
    MgS emission (dashed black line).}
  \label{fig:opticallythick}
\end{figure}
The feature found in CD-49~11554, IRAS~13416, RAFGL~618 and RAFGL~2688
is different from the others in the sense that it is broader and
flatter. This is especially true for RAFGL~618 and RAFGL~2688 where
the profile even shows a central depression. We investigate the effect
of optical depth on the emission profile. We model the effect of
optical depth with two simple limiting cases.
\begin{eqnarray}
  \label{eq:opticallythick1}
  I_{1}(\lambda) &= &B(\lambda,T_\mathrm{MgS}) \times (1-e^{-\rho\kappa_{\lambda}l}) ~\mathrm{and}\\
  \label{eq:opticallythick2}
  I_{2}(\lambda) &= &I_{0}(\lambda,T_\mathrm{MgS}) \times e^{-\rho\kappa_{\lambda}l},
\end{eqnarray}
where $T_\mathrm{MgS}$ is the temperature of the MgS, $B(\lambda,T)$
is the Planck function of temperature $T$, $\rho$ is the mass density
of MgS, $l$ is the column length and $I_{0}(\lambda,T)$ is MgS
emission of temperature $T$.  $I_{1}(\lambda)$ is the limiting case of
a column of MgS with a single temperature of $T_\mathrm{MgS}$.
$I_{2}(\lambda)$ represents the case of foreground absorption only;
i.e., MgS emission of temperature $T_\mathrm{MgS}$ obscured by a
column of MgS with negligible emission. In both cases the resulting
profile will be broader than the optically thin case.  For a long
column $I_{1}(\lambda)$ approaches $B(\lambda,T_\mathrm{MgS})$ while
$I_{2}(\lambda)$ becomes double peaked with a depression where
$\kappa_{\lambda}$ peaks. The resulting $I_{1}(\lambda)$ profile will
never show a central depression. We show the effects of incorporating
the optical depth in Fig.~\ref{fig:opticallythick}. Indeed, these
methods yield a broadened profile closer to what is observed. The
second method reproduces the central depression found in RAFGL~2688.
It is important to stress that the curves shown in
Fig.~\ref{fig:opticallythick} are not the result of a proper radiative
transfer modelling of the CS shell or an attempt to fit the observed
spectrum of the source. Nevertheless, they are able to explain the
general characteristics of the ``30'' $\mu$m profile in the deviant
sources well.

\subsection{PNe profiles}
\label{sec:pne-profiles}
\begin{figure}
  \includegraphics[width=8.8cm]{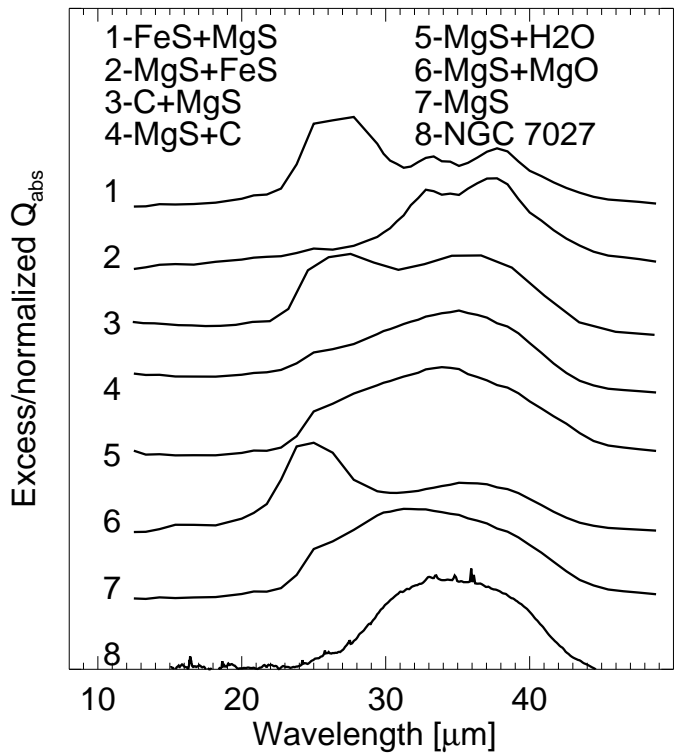}
  \caption{The absorptivity of MgS grains coated with
    various types of materials. The modelled grains are composed of 80
    per cent MgS and 20 per cent coating by volume. The models have
    been obtained by adding 5 ellipsoidal core-mantle grain of axes
    ratios (10:1:1),(3:1:1),(1:1:1),(3:3:1) and (10:10:1) of equal
    volume. For comparison we show the absorptivity of MgS in a CDE
    distribution (7) and the excess in NGC~7027 (8). We show the
    absorptivity of FeS (1) and amorphous carbon (5) grains with a MgS
    coating, both with core volume of 20 per cent of the grain. None
    of the modelled composite grains is able to explain the profile of
    the emission found in many PNe.}
  \label{fig:coatings}
\end{figure}
For 9 out of 13 PNe the observed ``30'' $\mu$m feature is much
narrower than in our model. Of the other four cases we discussed the
profile of RAFGL~618 in Sect.~\ref{sec:opticallythick}. It is
important to note that the remaining three cases (NGC 6790, IRAS~18240
and K 2-16) are the PNe with the highest continuum temperature among
the PNe in our sample.

In Fig.~\ref{fig:featurespectra}, we compare the shape of the ``30''
$\mu$m profile of NGC 7027 with the profiles due to differently shaped
MgS grains. As can be seen an oblate MgS grain with an axes ratio of
10:10:1 exhibits a ``30'' $\mu$m feature which peaks at the right
position. At present we don't know of a physical reason for a
preferred oblate grain shape in PNe, and a broader CDE shape
distribution in the C-stars and post-AGB objects (see also
Sect.~\ref{sec:planetarynebulae}).

The shape of a resonance is also influenced by the presence of a
coating.  MgS is very hygroscopic. Under conditions where oxygen is
available in the gas phase MgS can be oxidised and transformed into
MgO \citep{1985ApJ...290L..41N,1994ApJ...423L..71B}. It is possible
that the MgS is transformed as the central star of the PN heats up and
the UV radiation progressively dissociates the CO molecules yielding
gas phase oxygen.  This could lead to MgS grains which are coated by a
thin layer of MgO.  We have modelled such grains using the
electrostatic approximation following
\citet[][chap.~5]{BohrenHuffman}.  The result is shown in
Fig.~\ref{fig:coatings}, curve 6. As can be seen the ``30'' $\mu$m
resonance is split into two features due to the MgO coating. The
feature at the red wavelength is shifted to longer wavelengths
compared to the pure MgS resonance. However the main feature is on the
blue side of 25 $\mu$m towards the strong resonance at 18 $\mu$m in
the pure MgO material, in clear contrast with the observations.

We explore other possible coatings on MgS grains to test their ability
to explain the narrow feature observed in the PNe and the lack of
emission at 26 $\mu$m. We find that of the composite grains we tested
none give a satisfactory explanation. Mixtures of MgS and FeS have
been discussed in the literature to investigate the nature of the
``30'' $\mu$m feature \citep{1994ApJ...423L..71B, 2001A&A...368..497M,
  2000ESASP.456..329H}. Curves 1 and 2 in Fig.~\ref{fig:coatings} show
the result of embedding an FeS core in a mantle of MgS and embedding a
MgS core in a mantle of FeS, respectively. The latter compares most
favourably with the position of the feature in the PNe. However, the
substructure found in the spectrum of the composite grain around
33$-$37 $\mu$m is not found in the PNe spectra.

\citet{1999A&A...345L..39S} examine grains of amorphous carbon with a
mantle of MgS to compare with the 30 $\mu$m feature in two post-AGB
objects. We show simulated spectra of such grains and MgS grains
coated with amorphous carbon in Fig.~\ref{fig:coatings}, curves 3 and
4 respectively. Curve 3 clearly does not match the observed feature in
the PNe. As can been seen in curve 4 the MgS grains coated with
amorphous carbon absorb less at 26 $\mu$m than pure MgS and are
therefore a better spectral match to the ``30'' $\mu$m feature of the
PNe. However, the feature to continuum ratio in these grains is about
a factor 2.5 lower than in the pure MgS grains requiring a factor 2.5
more mass in the MgS component in order to explain the observed band
strength. Note also that such grains will still produce a weak feature
at 26 while in some PNe spectra we find no excess at that wavelength
at all.

Lastly, in curve 5 (Fig.~\ref{fig:coatings}) we show the effect of
water ice on the MgS grains. The effects on the optical properties of
a water ice coating are marginal and the profile cannot explain the
PNe observations. We conclude that of the composite materials we have
experimented with MgS grains coated with amorphous carbon give the
best spectral match.  However we find no composite grains that match
satisfactorily.

We stress that although our model does not reproduce the ``30'' $\mu$m
profile in the PNe in its width it is safe to assume that \emph{its
  carrier is MgS based}. These PNe are believed to be the evolutionary
descendants of the sources which exhibit the MgS feature.  The shift
in peak position compared to the post-AGB objects follows naturally
from an expanding and cooling shell. Also, the feature strength for
the PNe is similar to these found in the post-AGBs further
strengthening the physical link between the MgS in the C-stars and the
post-AGBs on one hand and the ``30'' $\mu$m feature in the PNe on the
other (see also Sect.~\ref{sec:correlations}).

\section{Correlations}
\label{sec:correlations}
\begin{figure}
  \includegraphics[width=8.8cm]{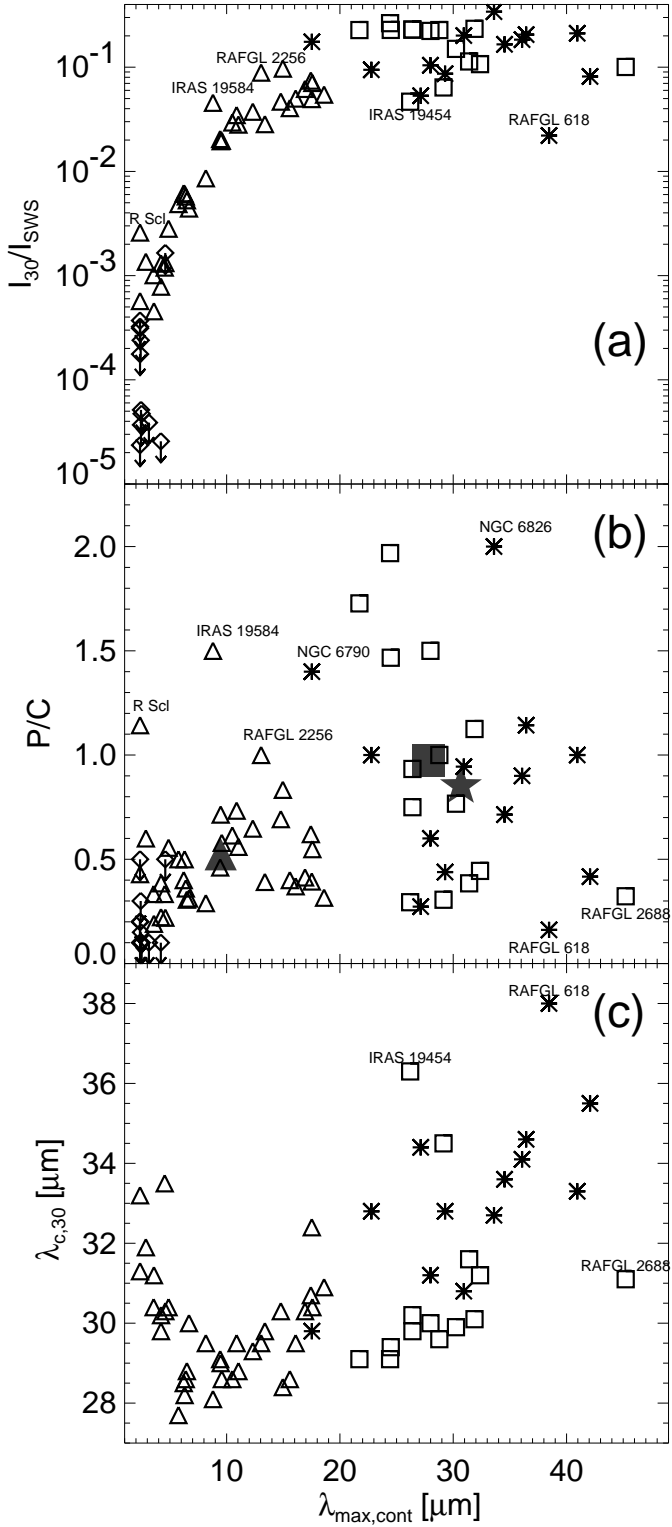}
  \caption{``30'' $\mu$m feature properties versus the peak wavelength
    of the continuum. The symbols are the same as in
    Fig.~\ref{fig:irascolours}. We show in panel (a) the ratio of the
    integrated flux in the ``30'' $\mu$m feature to the integrated
    flux in the SWS spectrum. In panel (b), we show the peak over
    continuum values. We also show the average values for the C-stars,
    the post-AGBs and the PNe. The non-detections are not taken into
    account in determining the mean values. The centroid wavelength of
    the ``30'' $\mu$m feature is shown in panel (c).}
  \label{fig:correlations}
\end{figure}
\begin{figure}
  \includegraphics[width=8.8cm]{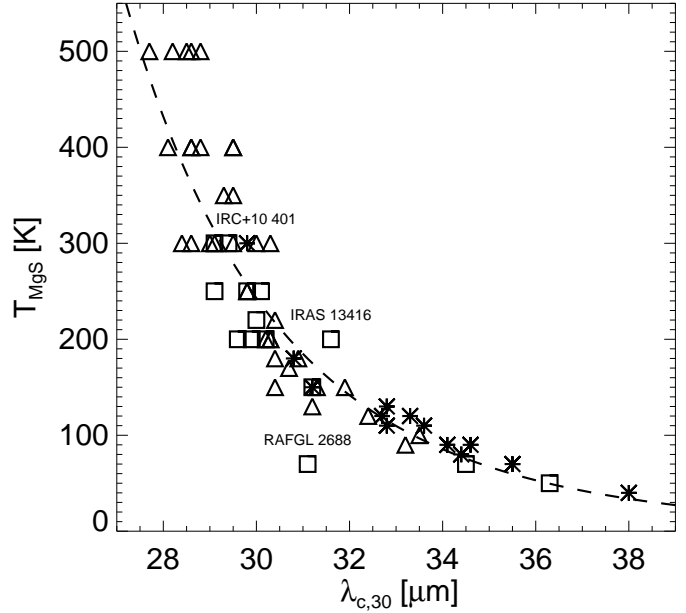}
  \caption{The centroid position of the ``30'' $\mu$m feature versus
    the MgS temperature in the model. The symbols are the same as in
    Fig.~\ref{fig:irascolours}. We show a power-law function fitted to
    the data with the dashed line.}
  \label{fig:centroidtemp}
\end{figure}
Using the large database of sources available we can study some of the
properties of the ``30'' $\mu$m feature statistically. We have found
it most convenient to characterise the sources by the temperature of
the fitted continuum.  We use the wavelength where the derived
continuum peaks ($\lambda_\mathrm{max,cont}$) as an indicator of the
continuum temperature.  The derived continuum temperature itself is
less well suited because of the systematic difference in the power law
index we find between classes of sources. The sources are rather
uniformly distributed over $\lambda_\mathrm{max,cont}$ as well.

First, we show in Fig.~\ref{fig:correlations}(a) the relation between
the $\lambda_\mathrm{max,cont}$ and the ratio of the integrated flux
in the ``30'' $\mu$m feature to the total flux in the SWS spectrum
(I$_{30}$/I$_\mathrm{SWS}$).  The C-stars demonstrate a clear increase
of I$_{30}$/I$_\mathrm{SWS}$ with decreasing continuum temperature.
The post-AGB objects emit systematically a larger fraction, of up to
25 per cent, in their ``30'' $\mu$m feature. The PNe emit a similar
fraction in the ``30'' $\mu$m feature as the post-AGB objects although
with a larger scatter. Notice that the sample contains a number of PNe
with warm dust indicative of young PNe. There are a few sources which
do not follow the general trend.  The C-stars, R~Scl, IRAS~19584 and
RAFGL~2256, exhibit an atypically strong ``30'' $\mu$m feature. These
latter two sources are further typified by very weak molecular
absorptions near 14 $\mu$m (see also Fig.~\ref{fig:spectra}). These
observed anomalies are indicative of deviating conditions in the
outflows of these sources, possibly a recently halted period of
efficient dust formation.  The post-AGB object IRAS~19454 has a very
weak and cold ``30'' $\mu$m feature.  RAFGL~618 has a weak feature due
to self-absorption (see Sect.~\ref{sec:opticallythick}).

The increasing strength of the ``30'' $\mu$m feature in the AGB stars
in not surprising. Since the emission is optically thin I$_{30}$ is
proportional to the amount of MgS. The low values of
I$_{30}$/I$_\mathrm{SWS}$ for the warmest C-stars reflects the fact
that there is little dust around these sources and most of the IR
radiation comes from the stellar photosphere. Cooler C-stars have more
dust and thus more MgS. The difference between the coolest C-stars and
the post-AGBs is more surprising.  The fact that post-AGBs emit a
larger fraction in the ``30'' $\mu$m feature is due to two effects.
First since the dust shell becomes optically thin in the visible some
fraction of the light is emitted at shorter wavelengths. Second, the
temperature of the MgS decreases less rapidly than the temperature of
the other dust components (see below).

It is clear that any dust component which produces 30 per cent of the
IR light has to be abundant. In order to quantify the (relative)
amounts of MgS present in the CS shells of these objects will require
radiative transfer modelling which is beyond the scope of this paper.
We can however in first approximation study the relative amounts of
MgS compared to the other cold dust components by studying the peak to
continuum ratio (P/C).  In Fig.~\ref{fig:correlations}(b), we show the
P/C versus the $\lambda_\mathrm{max,cont}$. The majority of the
sources lies within the 0.3$-$1.0 range in P/C. We indicate a few
clear outliers. R~Scl, IRAS~19584 and RAFGL~2256 have a very strong
``30'' $\mu$m feature indicating again that these sources have 'too
much' MgS for a normal C-star.  The PNe NGC~6790 and NGC~6826 have an
exceptionally strong MgS feature. Note that NGC~6790 also has a very
warm continuum, much like a post-AGB source or a very young PN. The
strong SiC band at 11 $\mu$m is consistent with this.  We also show
the averages for each of the classes of sources. The average P/C for
C-stars is 0.5, for post-AGB objects 1.0 and for the PNe it is 0.9.
The similar ratios for the post-AGB objects and the PNe suggests that
the carrier of the ``30'' $\mu$m feature in the PNe is indeed directly
related to the MgS feature in the post-AGBs.  Furthermore, the similar
ranges found for the post-AGB objects and the PNe argues against any
process which results in a destruction of the MgS grains during the PN
phase.

In Fig.~\ref{fig:centroidtemp}, we show the derived MgS temperature
versus the centroid position of the ``30'' $\mu$m feature. The two are
well correlated. For convenience, we have fitted a power-law function
(without physical meaning) to the relation.
\begin{equation}
  \label{eq:centroidtemperature}
  T_\mathrm{MgS} = 5.1\,10^{14}\,\left(\lambda_\mathrm{c,30}\right)^{-8.34}
\end{equation}
It is not surprising that $T_\mathrm{MgS}$ and $\lambda_\mathrm{c,30}$
are correlated since we have used the feature profile to estimate the
$T_\mathrm{MgS}$.  However, using Fig.~\ref{fig:centroidtemp} or
Eq.~(\ref{eq:centroidtemperature}) one can easily derive the MgS
temperature from a given observation. Also indicated in the figure are
IRAS~13416 and RAFGL~2688; as can be seen they fall outside the
correlation. This is due to the optical depth effects discussed above.

\begin{figure}
  \includegraphics[width=8.8cm]{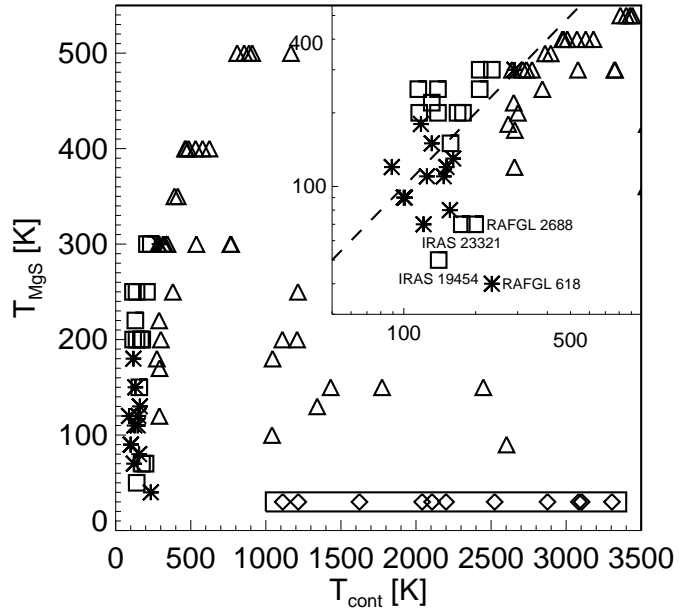}
  \caption{The derived MgS temperature versus the continuum
    temperature. The symbols are the same as in
    Fig.~\ref{fig:irascolours}. We show in the box in the lower right
    the continuum temperature of the sources without a ``30'' $\mu$m
    feature detected. The inset shows a blow up on a logarithmic scale
    up of the sources with a continuum temperature below 1000~K.}
  \label{fig:tmgstcont}
\end{figure}
Lastly, we show in Fig.~\ref{fig:tmgstcont} the relation we find
between the temperature of the continuum ($T_\mathrm{cont}$) and the
temperature of the MgS ($T_\mathrm{MgS}$). We find, tracing the
evolution from hot dust sources (C-stars) to the PNe, that the MgS
temperature decreases correspondingly. Surprisingly, we find for
warmest sources in the sample, with $T_\mathrm{cont}$ $>$ 1000~K, very
cold MgS. We propose two explanations for this phenomenon. First, it
may be due to the absorption properties of MgS. If MgS cannot
efficiently absorb the stellar light in the optical or near-IR part of
the spectrum the grains remain cold.  Alternatively, the MgS grains
may be located further away from the star (see below). In the inset,
we show a blow-up of the left side of the figure.  There is a clear
correlation between $T_\mathrm{cont}$ and $T_\mathrm{MgS}$. The MgS in
the post-AGB sources is systematically warmer than expected on the
basis of the C-stars. Apparently, in the process of becoming a
post-AGB object, when the dust shell becomes detached and moves away
from the star, the general dust cools down more rapidly than does the
MgS. Note that the difference in dust temperature between the coldest
C-stars and the warmest post-AGBs is seen most clearly in the
[12]-[25] colour or $\lambda_\mathrm{max,cont}$ but is also found in
the [25]-[60] colour (e.g.  Fig.~\ref{fig:irascolours}). The different
behaviour of the continuum and the MgS is nicely illustrated in
Fig.~\ref{fig:correlations}(c) where we show that while the continuum
becomes redder the position of the ``30'' $\mu$m feature does not
change. This effect is the cause for the discontinuity between the
C-stars and the post-AGB object seen in Fig.~\ref{fig:correlations}(a)
and (b). The physical cause for this must be due to the fact that MgS
is actually (partially) heated by mid-IR radiation. The mid-IR
radiation is much slower to respond to the termination of the AGB.
Conversely, this implies that mid-IR radiation is a less important
heating agent for the other dust components present.

\section{Discussion}
\label{sec:discussion}
\subsection{Carrier}
\label{sec:carrier}
The models using MgS we present do very well in explaining the profile
of the feature found in the C-stars and post-AGB objects. The model
does not do as well in explaining the profile of the feature found in
the PNe.  However, the smooth relation found between the [25]-[60]
colour and the $\lambda_\mathrm{c,30}$ shows a gradually changing
profile with dust temperature. This suggests one carrier or closely
related carriers.  Also the similar P/C values for the different
classes of objects is indicative of one carrier.

\subsection{The effect of model simplifications}
It is remarkable that our simple model works so well. We have applied
various simplifications in order to be able to study the whole sample
in a unified way. These simplifications necessarily lead to
differences between the models and the observations. Instead of
assuming a single temperature modified blackbody for the shape of the
continuum one can certainly obtain a more detailed fit by taking into
account the structure of the CS envelope and the opacities of the
different molecular and dust constituents. Our method works well
because the contributions of the other known other dust components
around 30 $\mu$m are well behaved, smooth and thus indeed form a
continuum.  Moreover, the shape of our modelled continuum between
20$-$50 $\mu$m is not a strong function of the emissivity index ($p$)
and varying the $p$-parameter or the temperature doesn't yield
qualitatively different continuum shapes or levels. We already pointed
out that in some cases our model fails at the longest wavelengths of
the SWS observations, which may be related to our method to estimate
the continuum.

We use a single temperature for the MgS to model its contribution. In
the very extended envelope of a C-star or in the nebulous environments
in the post-AGBs and PNe, the temperature of any dust component will
not be constant but decreases as a function of the distance to the
star. The fact that we still get good results using a single
temperature is a clear indication that the emission is optically thin
and that the density of the MgS falls off sharply with distance.  In
an optically thin environment we know that
$T_\mathrm{dust}\propto\sqrt{R}$, where R is the distance to the star.
If the density distribution drops with distance as R$^{-2}$ or
steeper, the contributions are weighted to the highest temperature
part of the envelope where $T_\mathrm{dust}$ and thus the source
function is highest. If the density distribution is flatter there is
relatively more dust far away than close by. In this case we will
observe MgS with a range of temperatures. Also sources that are not
optically thin will emit a feature broader than our single temperature
MgS model (see also Sect.~\ref{sec:opticallythick}).  We conclude that
in the majority of the sources the ``30'' $\mu$m emission is due to
optically thin emission which is dominated by the highest temperature
MgS closest to the star.

Recently \citet{2001ApJ...558L.129G} proposed HAC as the carrier of
the mid-IR emission features in C-stars, post-AGBs and PNe. HAC is a
very plausible dust component in those environments
\citep[e.g.][]{1981MNRAS.196..269D, 1987paha.proc..329G,
  1987ApJ...314..422B, 1999EM&P...80..179H}. However, the mid and
far-IR optical properties of HAC are dominated by a $\lambda^{-1}$
continuum \citep{1987paha.proc...63B}. \citet{2001ApJ...558L.129G}
identified some 13 weak spectral features in the range from 19$-$120
$\mu$m. The strongest of which occur at 21,27 and 57 $\mu$m. We
consider HAC a unlikely candidate for the circumstellar ``30'' $\mu$m
emission feature because the laboratory features are weak in contrast
to the astronomical data and because the far-IR features of HAC are
not observed in the astronomical spectra.

\subsection{Shape}
\label{sec:shape}
We find that a CDE shape distribution fits many observations well. A
CDE shape distribution is used to model a collection of irregularly
shaped grains. Irregularly shaped grains can for example result from
grain growth by agglomeration \citep{2000PhRvL..85.2426B}.

We find evidence for differences in the shape distributions between
sources. We have tested for correlations between the strength of the
26 $\mu$m excess (due to spherical MgS grain) and the CDE component
and other parameters like the mass-loss rate, the P/C, the continuum
temperature or the feature temperature. We find no clear correlations.
We do however note that we find little evidence for the 26 $\mu$m
excess in the hottest C-stars. Stars in our sample with a continuum
temperature above 1000~K do not exhibit the excess. Below 1000~K we
find both sources with and without the 26 $\mu$m excess. Note also
that the occurrence rate of the 26 $\mu$m excess in the sample of
post-AGB objects is high.  Because the emission in the post-AGB phase
may be dominated by the dust closest to the star and hence lost at the
tip of the AGB during a phase of heavy mass loss (at a rate much
higher than during the general AGB phase), this might suggest that the
shape distribution of the grains changes to become more spherical
towards the end of the AGB, possibly as a function of mass-loss rate.

\subsection{Planetary nebulae}
\label{sec:planetarynebulae}
The profile of the ``30'' $\mu$m feature we find in most PNe is
narrower than in our model. The largest discrepancy occurs around 26
$\mu$m. Our model fits are able to explain the shift in centroid
position observed in the PNe;  however, the models that explain the
band-shift best, over-predict the flux at 26 and 40 $\mu$m.

There are several observational properties that are important when
considering explanations for the observed discrepancies. First, there
is the similar values of P/C in the sample of post-AGBs and PNe
(Fig.~\ref{fig:correlations}b). This indicates that the carriers are
related and similar in abundance. Second, the smooth trend we find in
the centroid position of the ``30'' $\mu$m feature with the [25]-[60]
colour (Fig.~\ref{fig:peakshift}). The PNe profiles follow the general
trend. This indicates that the main effect for the peak shift
\emph{also in the PNe} is due to temperature. We also note that there
are three non-PNe sources in our sample with cold MgS (R~Scl,
IRAS~19454 and IRAS~23321). These sources are well fitted by our
model. This indicates that the narrow profiles are particular to
the PN environment.

We have explored possible MgS based heterogeneous grains and
variations in MgS grain shape. The heterogeneous grains we have
explored do not compare satisfactorily with the observed profile in
the PNe.  MgS grains coated with a layer of amorphous carbon provide a
somewhat better match in terms of the band shape however the contrast
of the feature with respect to the continuum is strongly reduced. The
reduced band strength is in contradiction with the observed peak over
continuum ratios.

We find that the emission from plate-like MgS grains appears similar
to the feature found in the PNe. In Sect.~\ref{sec:26-mum-excess} we
consider an extra contribution of spherical MgS grain to the ``30''
$\mu$m feature at 26 $\mu$m. The varying strength of this contribution
demonstrates that there are variations in the grains shape
distribution between sources. The question is: ``Is it possible that
the narrow ``30'' $\mu$m feature in the PNe is carried by flattened
MgS grains?''  This would either imply that the MgS grain shape
distribution in the PNe is heavily skewed to the plate-like grains or
that the oblate grains emit more efficiently.

The origin of these variations in the shape distribution are unclear.
They may reflect variations in the formation and destruction of
grains.  We do not know of a mechanism that drives towards oblate
grains shapes during the PN-phase or a mechanism that selectively
destroys spherical and prolate grains. Note, that the peak over
continuum value in the PNe is similar to those found in the post-AGB
objects arguing against a destruction of a part of the MgS grains. It
is important to remark that in the CDE distribution only $\sim$20 per
cent of the grains is oblate with an axis ratio of 1:${1}\over{5}$ or
more extreme, i.e.  with axes ratios of 1:X:Y, where Y is smaller than
${1}\over{5}$ and X can vary between $\sqrt{X}$ and 1. Such axes
ratios are required in order to shift the peak position to the
wavelength where the ``30'' $\mu$m feature in the PNe peaks and to
suppress the emission at 26 $\mu$m.

Alternatively, these inferred shape variations may result from a shape
dependent heating of grains. In the C-stars, likely, heating and
cooling occurs at IR wavelengths. Because the grains are small
compared to these wavelengths the temperature will be shape
independent. However, in the PNe heating occurs through absorption of
visible and UV radiation, which scales with the cross-section.  Hence
in that case dust temperatures will be shape dependent. Indeed
flattened grains have on average a larger geometric cross-section
compared to their volume that spherical or prolate grains.

As noted before, the PNe in the sample with the warmest dust continuum
are well fitted by our model. The model fails in the PNe sample at MgS
temperatures below $\sim$90~K. Perhaps, the observed discrepancies are
due to changes in the optical properties of pure MgS at low
temperatures. No laboratory measurements of MgS at these temperatures
are published. FeO has the same lattice structure as MgS and
measurements of FeO at different temperatures (10~K, 100~K, 200~K and
300~K) are available \citep{1997A&A...327..743H}. We have used FeO as
an analogue to examine the effect of temperature. The trend with lower
temperature is for the FeO resonance at 20 $\mu$m to become narrower
and stronger with respect to the continuum. There is a small shift in
the peak position. Comparing the 300~K measurement to 10~K the latter
resonance peaks $\sim$0.1 $\mu$m more to the blue for spherical
grains. Like MgS, FeO is also very sensitive to the shape of the
grains. Using a CDE shape distribution the resonance of the 10~K
sample lies $\sim$1 $\mu$m on the blue side of the 300~K resonance. If
we translate this behaviour to the MgS data and the ``30'' $\mu$m
feature in the PNe it worsens the situation since we would expect a
stronger contribution at shorter wavelengths exactly where our model
already over-predicts.  As an alternative, one might speculate that a
change in lattice structure at low temperatures may occur resulting in
different optical properties. It is interesting to note that
  \citet{Berthold1964} reports that MgS can condense in two lattice
  structures: cubic and hexagonal. This author finds that hexagonal
  MgS exhibits a narrower mid-IR resonance than cubic and amorphous
  MgS.

We conclude that, although it is possible to find MgS based candidate
materials that provide a better spectral match to the observed ``30''
$\mu$m feature in the PNe no explanation for the ``30'' $\mu$m
profiles in the PNe is completely satisfactory at this time.

\subsection{MgS Temperature}
\label{sec:mgstemp}
We have found that the ``30'' $\mu$m feature is very well explained
with MgS provided that the continuum temperature and the MgS can be
substantially different from each other (see also
Fig.~\ref{fig:tmgstcont}). Moreover, the temperature difference
changes between classes of sources. In the C-stars we find that MgS is
colder than the continuum while in the post-AGB sample the MgS grains
are warmer. The temperature difference and the fact that the continuum
temperature varies relative to the MgS temperature implies that the
MgS grain are not in thermal contact with the amorphous carbons grains
that carry the continuum. Thus, MgS is condensed in these environments
into separate grains.

The relatively warm hot MgS that we observe in the post-AGB sample is
certainly a phenomenon that deserves further study. As pointed out
above, this must be due to a difference in the heating properties of
MgS relative to the other dust constituents. We suggest that due to
the strong resonance in the mid-IR MgS is (partially) heated by IR
radiation. This may cause a temperature difference as observed. It
requires measured optical constants for MgS over a much wider
wavelength range than now available and radiative transfer modelling
to understand the detailed temperature behaviour of the MgS grains.

We have shown in Fig.~\ref{fig:tmgstcont} that the warmest C-stars
that exhibit the ``30'' $\mu$m feature have cold MgS. We consider two
possible explanations for this behaviour. First, MgS can be cold
because it is unable to absorb the radiation from the central source
efficiently. This would require that the MgS grains are very
transparent in the visible and near-IR where these warm C-stars emit
most of their radiation. In this case a \emph{cooler} star will emit
more radiation at wavelengths where MgS can absorb and as a result the
MgS grains will be \emph{warmer}. This explanation is consistent with
our notion that MgS is (partially) heated by IR radiation.

A second intriguing possibility is that the MgS around these sources
is located far away from the star. R~Scl, U~Cam and W~Ori are known to
have a detached dust shell \citep{1993A&A...276..367Z}.  These shells
are formed during an earlier phase when the mass-loss rate was higher
than the current mass-loss rate \citep{1988A&A...196..173W}. If during
this phase the AGB star was already a C-star, MgS could have condensed
in the outflow and be present in the detached shell. We have sketched
in Fig.~\ref{fig:irascolours} the evolution of a star which suffers a
brief period of enhanced mass loss. First, the star becomes redder
during the phase of high mass loss. When the shell becomes detached
the 12 $\mu$m flux quickly drops while the inner edge of the dust
shell moves away from the star and the warmest dust is rapidly lost.
The star moves back to the locus of the warmest C-stars but with an
excess of cold dust. The excess of cold dust is observable as a 60
$\mu$m excess. The sources mentioned above indeed show the 60 $\mu$m
excess.  We note that some stars with 60 $\mu$m excess do not show the
``30'' $\mu$m feature.

To distinguish between these two possibilities requires radiative
transfer modelling and a more detailed investigation of these
particular sources which is beyond the scope of this paper. The
location and temperature of the MgS in the warmest C-stars will be
further discussed in (Hony et al., in prep.).

\subsection{MgS in the ISM}
\label{sec:ism}
The behaviour of the MgS we find in this study has implications for
our understanding of the fate of MgS that is produced during the
C-star phase. Previously, the existence of MgS in the ISM has been
excluded because of the lack of spectral signature at 26 $\mu$m. We
have shown that at low dust temperatures no emission signature is
expected at 26 $\mu$m. For the dust temperatures in the ISM we would
expect a smooth feature at wavelengths longer than 35 $\mu$m. In this
respect it is interesting to note that the peak to continuum (P/C)
values in the post-AGB and PNe sample are similar. We find no evidence
for a rapid destruction by the UV radiation field in the PN. This
shows that the MgS will indeed be injected into the ISM. Of course, in
the ISM the contrast will be much less because the injected material
will be diluted with other star dust and general ISM material.
Nevertheless a search for the ``30'' $\mu$m feature in the ISM may be
very worthwhile. In this respect it is important to note that
\citet{1997ApJ...483..798C} have discussed a broad emission feature
they observed along several lines of sight towards the galactic
centre. The feature they observed is similar to the ``30'' $\mu$m
feature observed in the C-rich post-AGB objects.

\section{Summary and concluding remarks}
We present a large sample of 63 ISO/SWS spectra of carbon-rich evolved
objects that exhibit a ``30'' $\mu$m feature. The feature is detected
along the complete evolutionary track from low mass loss C-stars and
extreme C-stars via post AGB objects to PNe.

We present a simple approach to determine the continuum. We study the
profile of the ``30'' $\mu$m feature after continuum subtraction. We
find large systematic variations in the appearance of the ``30''
$\mu$m feature. We firmly identify the carrier of the feature with MgS
in a CDE shape distribution. The profile of the observed feature is
consistent with MgS provided that the temperature of the MgS can be
different from the bulk of the dust. This approach allows us to model,
in a unified way, the profiles of the ``30'' $\mu$m feature in a wide
range of objects even when the feature can appear extremely different.

We find an additional component at 26 $\mu$m in $\sim$25 sources. We
argue that this component is due to differences in the distribution
over shapes of the MgS grains; specifically, it requires a
distribution more weighted to spherical MgS grains.  We find no clear
correlations of this excess with other properties of the sources. The
self-absorbed ``30'' $\mu$m features of RAFGL~618 and RAFGL~2688
reflect the high optical depth in the ``30'' $\mu$m feature in these
sources.

We find that the typical profile of the ``30'' $\mu$m feature in the
PNe is narrower than predicted by our model. We consider several
possible explanations. We find that flattened MgS grains provide a
better spectral match to the ``30'' $\mu$m feature in the PNe. However
it is presently unclear why in the PNe environment the ``30'' $\mu$m
feature is dominated by flattened grains.

We find that the temperature of the MgS grains is different from the
bulk of the dust. Therefore, they cannot be in thermal contact with
the other dust species but must exist as separate grains. In the
C-stars MgS is colder than the other dust while in the post-AGB
objects MgS is warmer. Likely, this is because the MgS is efficiently
heated by mid-IR radiation which is less important in the C-stars.

The behaviour of the temperature of the MgS for the hottest C-stars is
enigmatic. The hottest sources have very cold MgS grains. We propose
two explanations for this phenomenon.  First, the ``30'' $\mu$m
feature in these sources may be due to a previous mass-loss phase and
the MgS is thus located far from the source and cold.  Second, the MgS
grains absorb very inefficiently in the optical and near-IR and
therefore the hottest C-stars do not heat the MgS grains well.

We examine the feature over continuum ratio to study the relative
proportion of MgS to the other dust components. We find no evidence
for rapid destruction of MgS grains during the PNe phase and possibly
MgS grains may survive to be incorporated into the ISM.

We would like to emphasise the need for further measurements of the
optical properties of astrophysically relevant materials at \emph{all
  relevant} wavelengths. This study has put forwards some very
interesting questions in relation to the absorption properties of MgS
in the optical and near-IR and also concerning low temperature
effects. Unfortunately, no published laboratory measurements are
available to test some of our scenarios and conclusions.

\begin{acknowledgements}
  SH and LBFMW acknowledge financial support from NWO \emph{Pionier}
  grant 616-78-333. This research has made use of the SIMBAD database,
  operated at CDS, Strasbourg, France.  This research has made use of
  NASA's Astrophysics Data System Bibliographic Services. IA$^3$ is a
  joint development of the SWS consortium.  Contributing institutes
  are SRON, MPE, KUL and the ESA Astrophysics Division.
\end{acknowledgements}

\bibliographystyle{aa}
\bibliography{articles}

\begin{thebibliography}{49}
\expandafter\ifx\csname natexlab\endcsname\relax\def\natexlab#1{#1}\fi

\bibitem[{{Begemann} {et~al.}(1994){Begemann}, {Dorschner}, {Henning},
  {Mutschke}, \& {Thamm}}]{1994ApJ...423L..71B}
{Begemann}, B., {Dorschner}, J., {Henning}, T., {Mutschke}, H., \& {Thamm}, E.
  1994, \apjl, 423, L71

\bibitem[{{Berthold}(1964)}]{Berthold1964}
{Berthold}, G. 1964, Zs. Phys., 181, 333

\bibitem[{{Blum} {et~al.}(2000){Blum}, {Wurm}, {Kempf}, {Poppe}, {Klahr},
  {Kozasa}, {Rott}, {Henning}, {Dorschner}, {Schr{\" a}pler}, {Keller},
  {Markiewicz}, {Mann}, {Gustafson}, {Giovane}, {Neuhaus}, {Fechtig}, {Gr{\"
  u}n}, {Feuerbacher}, {Kochan}, {Ratke}, {El Goresy}, {Morfill},
  {Weidenschilling}, {Schwehm}, {Metzler}, \& {Ip}}]{2000PhRvL..85.2426B}
{Blum}, J., {Wurm}, G., {Kempf}, S., {et~al.} 2000, Physical Review Letters,
  85, 2426

\bibitem[{{Bohren} \& {Huffman}(1983)}]{BohrenHuffman}
{Bohren}, C.~F. \& {Huffman}, D.~R. 1983, Absorption and scattering of light by
  small particles (New York: Wiley, 1983)

\bibitem[{{Borghesi} {et~al.}(1987){Borghesi}, {Bussoletti}, \&
  {Colangeli}}]{1987ApJ...314..422B}
{Borghesi}, A., {Bussoletti}, E., \& {Colangeli}, L. 1987, \apj, 314, 422

\bibitem[{{Bussoletti} {et~al.}(1987){Bussoletti}, {Colangeli}, \&
  {Borghesi}}]{1987paha.proc...63B}
{Bussoletti}, E., {Colangeli}, L., \& {Borghesi}, A. 1987, in NATO ASIC Proc.
  191: Polycyclic Aromatic Hydrocarbons and Astrophysics, 63

\bibitem[{{Cernicharo} {et~al.}(2001){Cernicharo}, {Heras}, {Pardo}, {Tielens},
  {Gu{\' e}lin}, {Dartois}, {Neri}, \& {Waters}}]{2001ApJ...546L.127C}
{Cernicharo}, J.~., {Heras}, A.~M., {Pardo}, J.~R., {et~al.} 2001, \apjl, 546,
  L127

\bibitem[{{Chan} {et~al.}(1997){Chan}, {Moseley}, {Casey}, {Harrington},
  {Dwek}, {Loewenstein}, {Varosi}, \& {Glaccum}}]{1997ApJ...483..798C}
{Chan}, K., {Moseley}, S.~H., {Casey}, S., {et~al.} 1997, \apj, 483, 798

\bibitem[{{Cox}(1993)}]{1993ais..conf..163C}
{Cox}, P. 1993, in ASP Conf. Ser. 41: Astronomical Infrared Spectroscopy:
  Future Observational Directions, 163

\bibitem[{{de Graauw} {et~al.}(1996){de Graauw}, {Haser}, {Beintema},
  {et~al.}}]{1996A&A...315L..49D}
{de Graauw}, T., {Haser}, L.~N., {Beintema}, D.~A., {et~al.} 1996, \aap, 315,
  L49

\bibitem[{{Duley}(2000)}]{2000ApJ...528..841D}
{Duley}, W.~W. 2000, \apj, 528, 841

\bibitem[{{Duley} \& {Williams}(1981)}]{1981MNRAS.196..269D}
{Duley}, W.~W. \& {Williams}, D.~A. 1981, \mnras, 196, 269

\bibitem[{{Forrest} {et~al.}(1981){Forrest}, {Houck}, \&
  {McCarthy}}]{1981ApJ...248..195F}
{Forrest}, W.~J., {Houck}, J.~R., \& {McCarthy}, J.~F. 1981, \apj, 248, 195

\bibitem[{{Goebel}(1987)}]{1987paha.proc..329G}
{Goebel}, J.~H. 1987, in NATO ASIC Proc. 191: Polycyclic Aromatic Hydrocarbons
  and Astrophysics, 329

\bibitem[{{Goebel} \& {Moseley}(1985)}]{1985ApJ...290L..35G}
{Goebel}, J.~H. \& {Moseley}, S.~H. 1985, \apjl, 290, L35

\bibitem[{{Grishko} {et~al.}(2001){Grishko}, {Tereszchuk}, {Duley}, \&
  {Bernath}}]{2001ApJ...558L.129G}
{Grishko}, V.~I., {Tereszchuk}, K., {Duley}, W.~W., \& {Bernath}, P. 2001,
  \apjl, 558, L129

\bibitem[{{Guillois} {et~al.}(1996){Guillois}, {Nenner}, {Papoular}, \&
  {Reynaud}}]{1996ApJ...464..810G}
{Guillois}, O., {Nenner}, I., {Papoular}, R., \& {Reynaud}, C. 1996, \apj, 464,
  810

\bibitem[{{Henning}(2000)}]{2000ESASP.456..329H}
{Henning}, T. 2000, ISO beyond the peaks: The 2nd ISO workshop on analytical
  spectroscopy.~Eds.~A.~Salama, M.F.Kessler, K.~Leech \& B.~Schulz.~ESA-SP
  456., 456, 329

\bibitem[{{Henning} \& {Mutschke}(1997)}]{1997A&A...327..743H}
{Henning}, T. \& {Mutschke}, H. 1997, \aap, 327, 743

\bibitem[{{Henning} \& {Schnaiter}(1999)}]{1999EM&P...80..179H}
{Henning}, T. \& {Schnaiter}, M. 1999, Earth Moon and Planets, 80, 179

\bibitem[{{Hony} {et~al.}(2002){Hony}, {Bouwman}, {Keller}, \&
  {Waters}}]{FeS_PNe_Hony}
{Hony}, S., {Bouwman}, J.~B., {Keller}, L.~P., \& {Waters}, L.~B.~F.~M. 2002,
  \aap, submitted

\bibitem[{{Hony} {et~al.}(2001){Hony}, {Waters}, \&
  {Tielens}}]{2001A&A...378L..41H}
{Hony}, S., {Waters}, L.~B.~F.~M., \& {Tielens}, A.~G.~G.~M. 2001, \aap, 378,
  L41

\bibitem[{{Hrivnak} {et~al.}(2000){Hrivnak}, {Volk}, \&
  {Kwok}}]{2000ApJ...535..275H}
{Hrivnak}, B.~J., {Volk}, K., \& {Kwok}, S. 2000, \apj, 535, 275

\bibitem[{{Jiang} {et~al.}(1999){Jiang}, {Szczerba}, \&
  {Deguchi}}]{1999A&A...344..918J}
{Jiang}, B.~W., {Szczerba}, R., \& {Deguchi}, S. 1999, \aap, 344, 918

\bibitem[{{Kaler} \& {Jacoby}(1991)}]{1991ApJ...372..215K}
{Kaler}, J.~B. \& {Jacoby}, G.~H. 1991, \apj, 372, 215

\bibitem[{{Keller} {et~al.}(2002)}]{keller_nature}
{Keller}, L. {et~al.} 2002, \nat, in press

\bibitem[{{Kessler} {et~al.}(1996){Kessler}, {Steinz}, {Anderegg},
  {et~al.}}]{1996A&A...315L..27K}
{Kessler}, M.~F., {Steinz}, J.~A., {Anderegg}, M.~E., {et~al.} 1996, \aap, 315,
  L27

\bibitem[{{Kwok} {et~al.}(2001){Kwok}, {Volk}, \&
  {Bernath}}]{2001ApJ...554L..87K}
{Kwok}, S., {Volk}, K., \& {Bernath}, P. 2001, \apjl, 554, L87

\bibitem[{{Latter} {et~al.}(2000){Latter}, {Dayal}, {Bieging}, {Meakin},
  {Hora}, {Kelly}, \& {Tielens}}]{2000ApJ...539..783L}
{Latter}, W.~B., {Dayal}, A., {Bieging}, J.~H., {et~al.} 2000, \apj, 539, 783

\bibitem[{{Lattimer} {et~al.}(1978){Lattimer}, {Schramm}, \&
  {Grossman}}]{1978ApJ...219..230L}
{Lattimer}, J.~M., {Schramm}, D.~N., \& {Grossman}, L. 1978, \apj, 219, 230

\bibitem[{{Lodders} \& {Fegley}(1999)}]{1999IAUS..191..279L}
{Lodders}, K. \& {Fegley}, B. 1999, in IAU Symp. 191: Asymptotic Giant Branch
  Stars, Vol. 191, 279

\bibitem[{{Mendez} {et~al.}(1992){Mendez}, {Kudritzki}, \&
  {Herrero}}]{1992A&A...260..329M}
{Mendez}, R.~H., {Kudritzki}, R.~P., \& {Herrero}, A. 1992, \aap, 260, 329

\bibitem[{{Men'shchikov} {et~al.}(2001){Men'shchikov}, {Balega}, {Bl{\"
  o}cker}, {Osterbart}, \& {Weigelt}}]{2001A&A...368..497M}
{Men'shchikov}, A.~B., {Balega}, Y., {Bl{\" o}cker}, T., {Osterbart}, R., \&
  {Weigelt}, G. 2001, \aap, 368, 497

\bibitem[{{Nuth} {et~al.}(1985){Nuth}, {Moseley}, {Silverberg}, {Goebel}, \&
  {Moore}}]{1985ApJ...290L..41N}
{Nuth}, J.~A., {Moseley}, S.~H., {Silverberg}, R.~F., {Goebel}, J.~H., \&
  {Moore}, W.~J. 1985, \apjl, 290, L41

\bibitem[{{Omont}(1993)}]{1993ais..conf...87O}
{Omont}, A. 1993, in ASP Conf. Ser. 41: Astronomical Infrared Spectroscopy:
  Future Observational Directions, 87

\bibitem[{{Omont} {et~al.}(1995){Omont}, {Moseley}, {Cox}, {Glaccum}, {Casey},
  {Forveille}, {Chan}, {Szczerba}, {Loewenstein}, {Harvey}, \&
  {Kwok}}]{1995ApJ...454..819O}
{Omont}, A., {Moseley}, S.~H., {Cox}, P., {et~al.} 1995, \apj, 454, 819

\bibitem[{{Papoular}(2000)}]{2000A&A...362L...9P}
{Papoular}, R. 2000, \aap, 362, L9

\bibitem[{{Perinotto}(1991)}]{1991ApJS...76..687P}
{Perinotto}, M. 1991, \apjs, 76, 687

\bibitem[{{Quigley} \& {Bruhweiler}(1995)}]{1995AAS...187.8008Q}
{Quigley}, M.~F. \& {Bruhweiler}, F.~C. 1995, in American Astronomical Society
  Meeting, Vol. 187, 8008

\bibitem[{{Szczerba} {et~al.}(1999){Szczerba}, {Henning}, {Volk}, {Kwok}, \&
  {Cox}}]{1999A&A...345L..39S}
{Szczerba}, R., {Henning}, T., {Volk}, K., {Kwok}, S., \& {Cox}, P. 1999, \aap,
  345, L39

\bibitem[{{van der Veen} \& {Habing}(1988)}]{1988A&A...194..125V}
{van der Veen}, W.~E.~C.~J. \& {Habing}, H.~J. 1988, \aap, 194, 125

\bibitem[{{Volk} {et~al.}(2002){Volk}, {Kwok}, {Hrivnak}, \&
  {Szczerba}}]{Volk2002}
{Volk}, K., {Kwok}, S., {Hrivnak}, B.~J., \& {Szczerba}, R. 2002, \apj, 567,
  412

\bibitem[{{Volk} {et~al.}(2000){Volk}, {Xiong}, \&
  {Kwok}}]{2000ApJ...530..408V}
{Volk}, K., {Xiong}, G., \& {Kwok}, S. 2000, \apj, 530, 408

\bibitem[{{von Helden} {et~al.}(2000){von Helden}, {Tielens}, {van
  Heijnsbergen}, {Duncan}, {Hony}, {Waters}, \& {Meijer}}]{2000Helden}
{von Helden}, G., {Tielens}, A. G. G.~M., {van Heijnsbergen}, D., {et~al.}
  2000, Science, 288, 313

\bibitem[{{Waters} {et~al.}(2000){Waters}, {Molster}, {Hony}, {Kemper},
  {Yamamura}, {de Jong}, {Tielens}, \& {Waelkens}}]{2000tesa.conf....3W}
{Waters}, L.~B.~F.~M., {Molster}, F.~J., {Hony}, S., {et~al.} 2000, ASP
  Conf.~Ser.~196: Thermal Emission Spectroscopy and Analysis of Dust, Disks,
  and Regoliths, 3

\bibitem[{{Willems} \& {de Jong}(1988)}]{1988A&A...196..173W}
{Willems}, F.~J. \& {de Jong}, T. 1988, \aap, 196, 173

\bibitem[{{Yamamura} {et~al.}(1998){Yamamura}, {De Jong}, {Justtanont}, {Cami},
  \& {Waters}}]{1998Ap&SS.255..351Y}
{Yamamura}, I., {De Jong}, T., {Justtanont}, K., {Cami}, J., \& {Waters}, L. B.
  F.~M. 1998, \apss, 255, 351

\bibitem[{{Young} {et~al.}(1993){Young}, {Phillips}, \&
  {Knapp}}]{1993ApJS...86..517Y}
{Young}, K., {Phillips}, T.~G., \& {Knapp}, G.~R. 1993, \apjs, 86, 517

\bibitem[{{Zuckerman}(1993)}]{1993A&A...276..367Z}
{Zuckerman}, B. 1993, \aap, 276, 367

\end{thebibliography}

\appendix
\section{Model Fits}
\begin{figure*}
  \includegraphics[width=17cm]{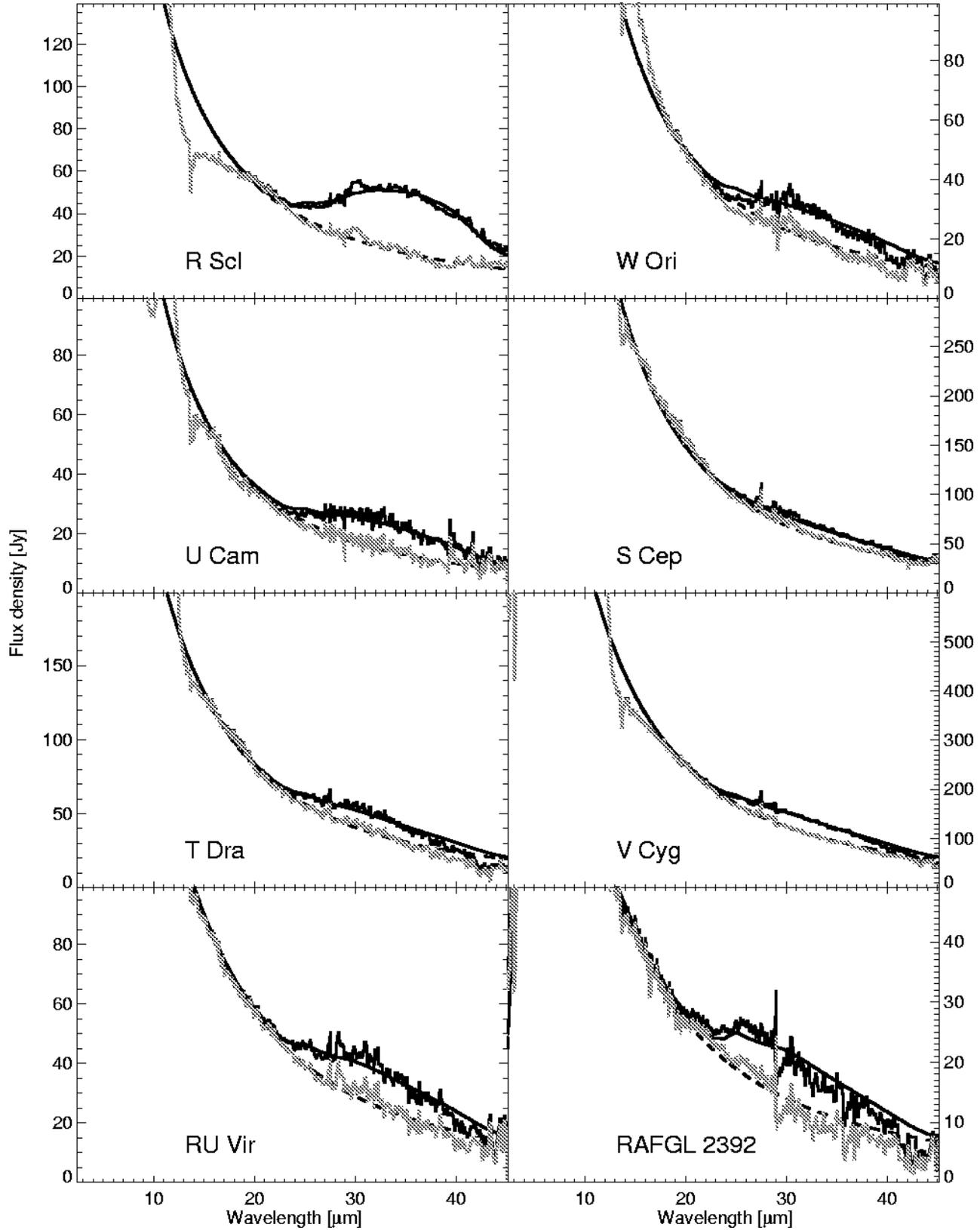}
  \caption{``30'' $\mu$m feature modelled using MgS optical properties.}
  \label{fig:fits}
\end{figure*}
\addtocounter{figure}{-1}
\begin{figure*}
  \includegraphics[width=17cm]{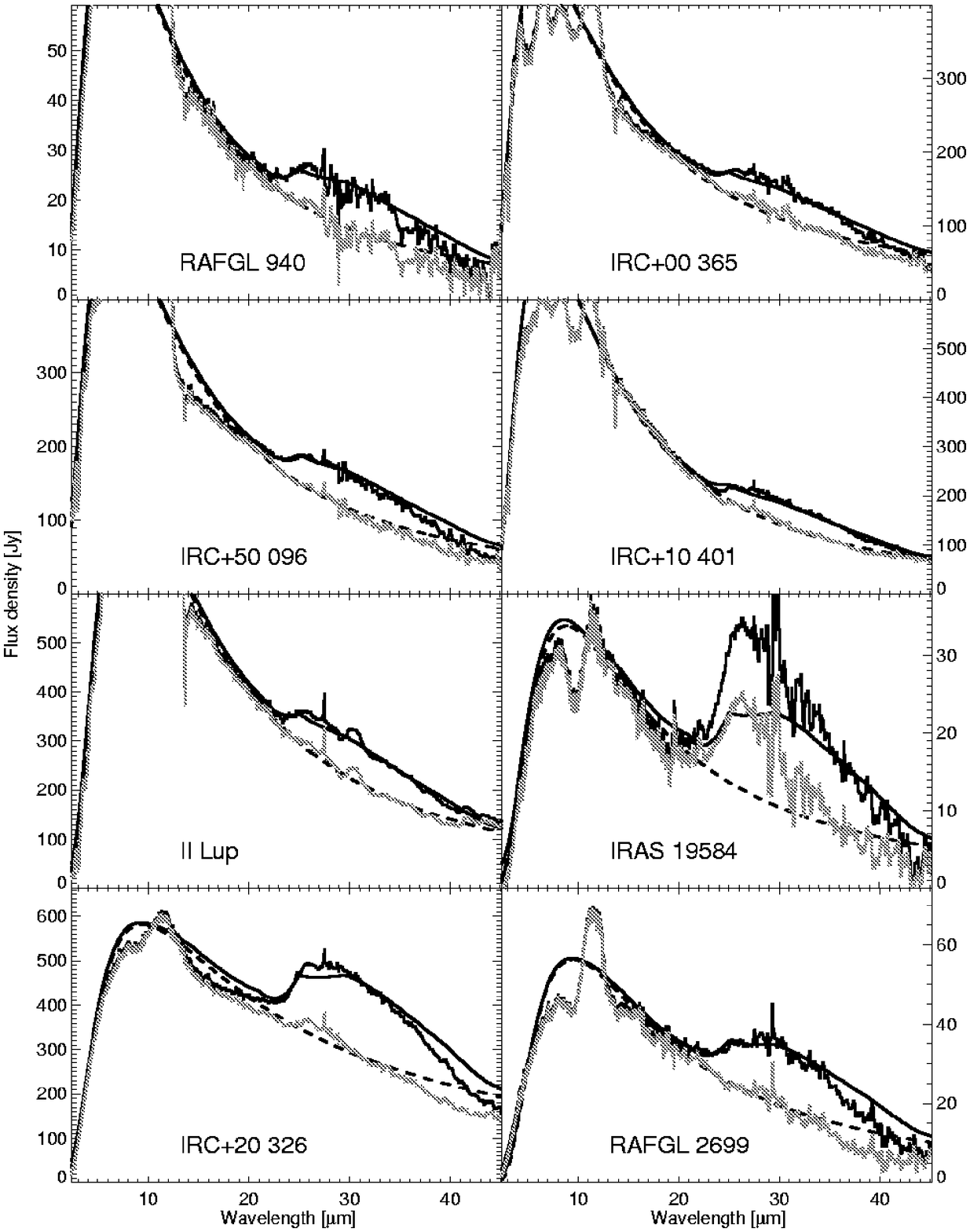}
  \caption{(continued).}
\end{figure*}
\addtocounter{figure}{-1}
\begin{figure*}
  \includegraphics[width=17cm]{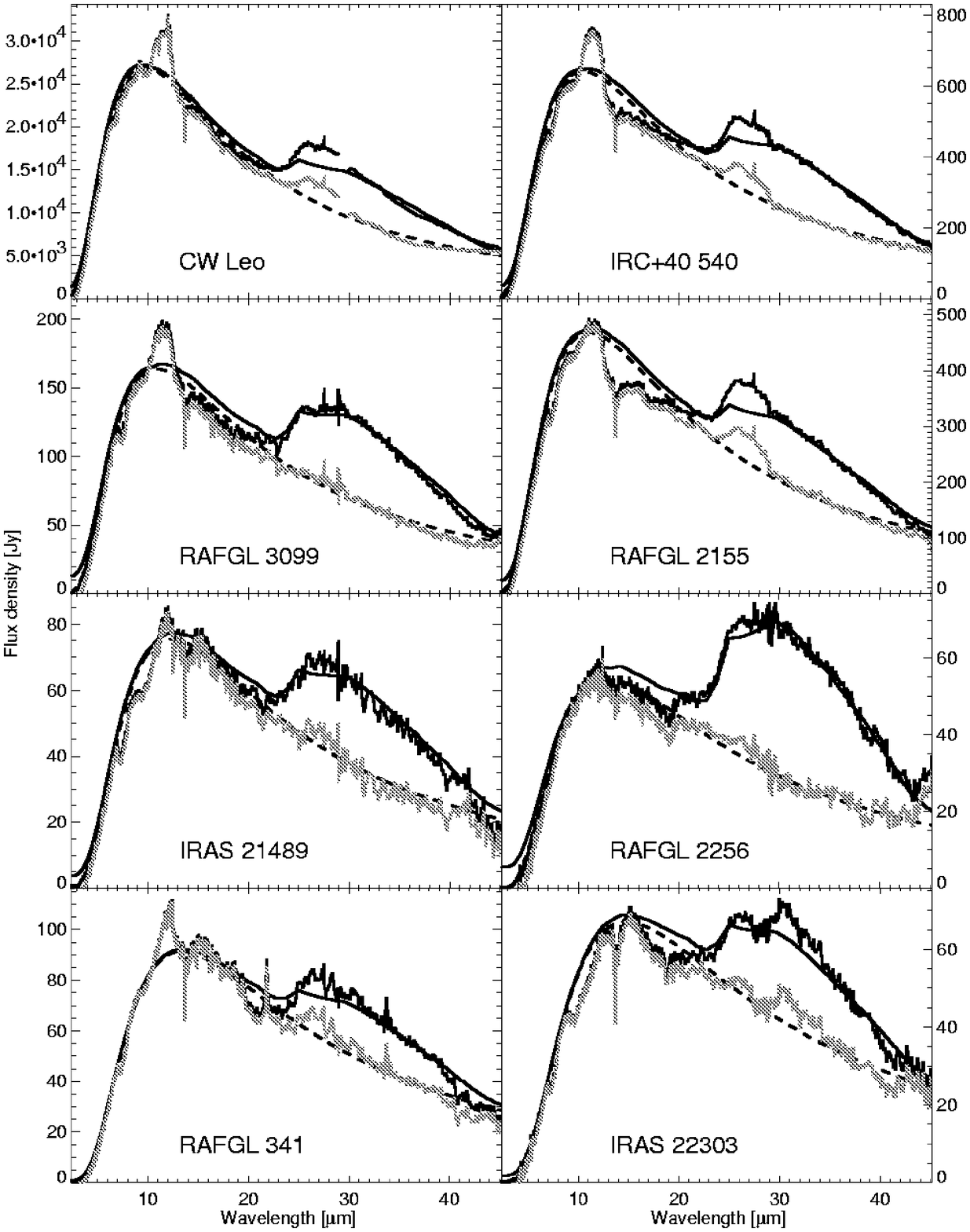}
  \caption{(continued).}
\end{figure*}
\addtocounter{figure}{-1}
\begin{figure*}
  \includegraphics[width=17cm]{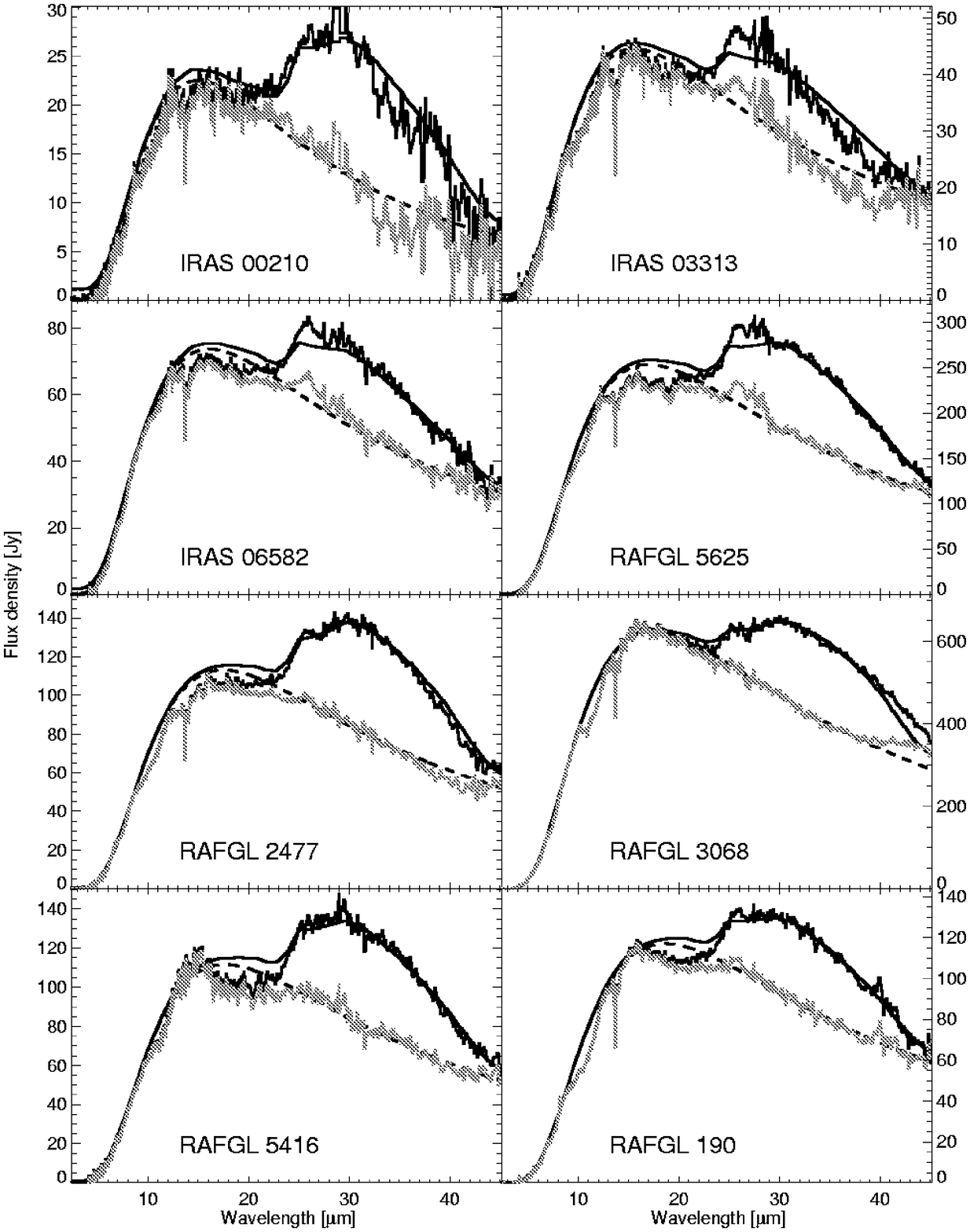}
  \caption{(continued).}
\end{figure*}
\addtocounter{figure}{-1}
\begin{figure*}
  \includegraphics[width=17cm]{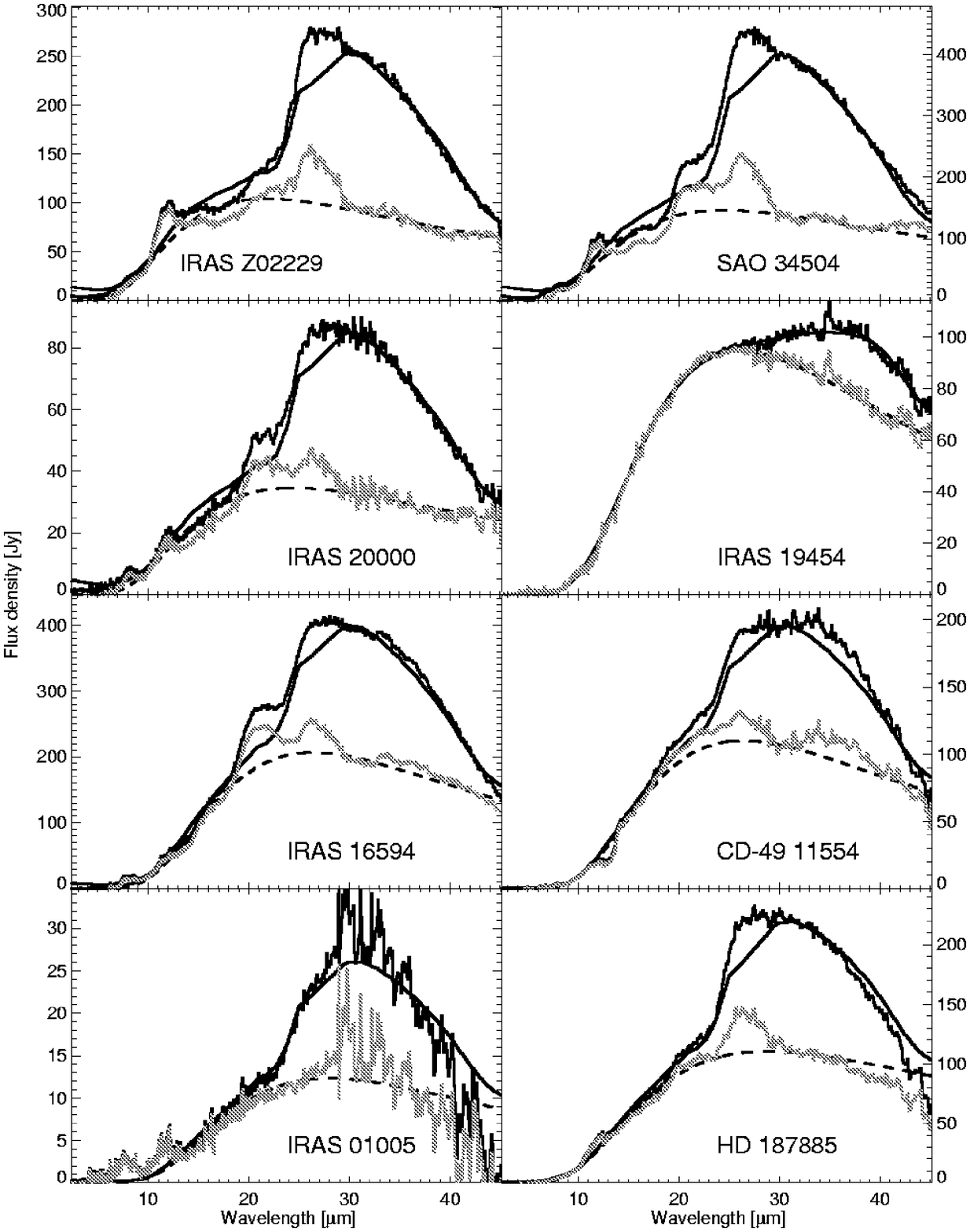}
  \caption{(continued).}
\end{figure*}
\addtocounter{figure}{-1}
\begin{figure*}
  \includegraphics[width=17cm]{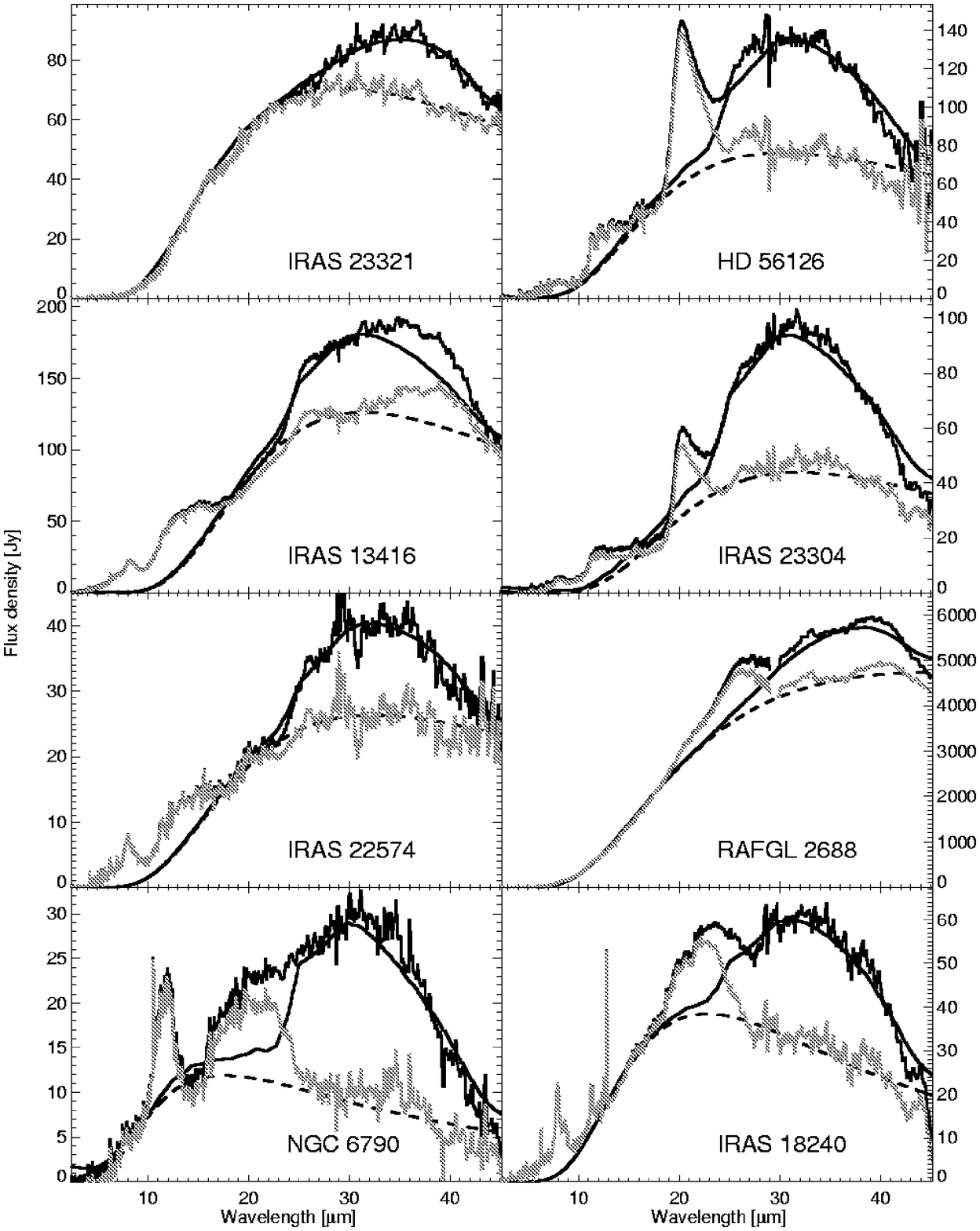}
  \caption{(continued).}
\end{figure*}
\addtocounter{figure}{-1}
\begin{figure*}
  \includegraphics[width=17cm]{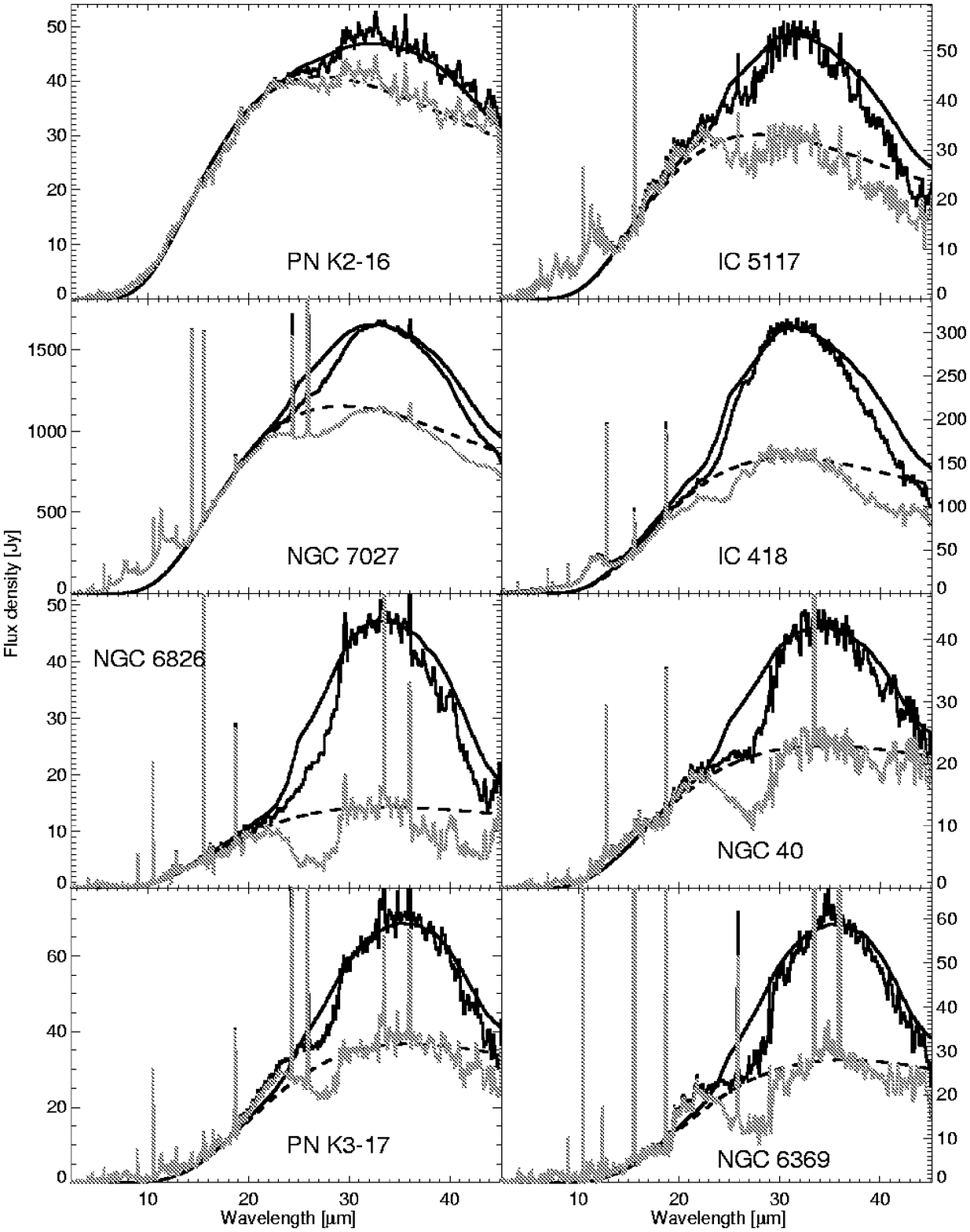}
  \caption{(continued).}
\end{figure*}
\addtocounter{figure}{-1}
\begin{figure*}
  \includegraphics[width=17cm]{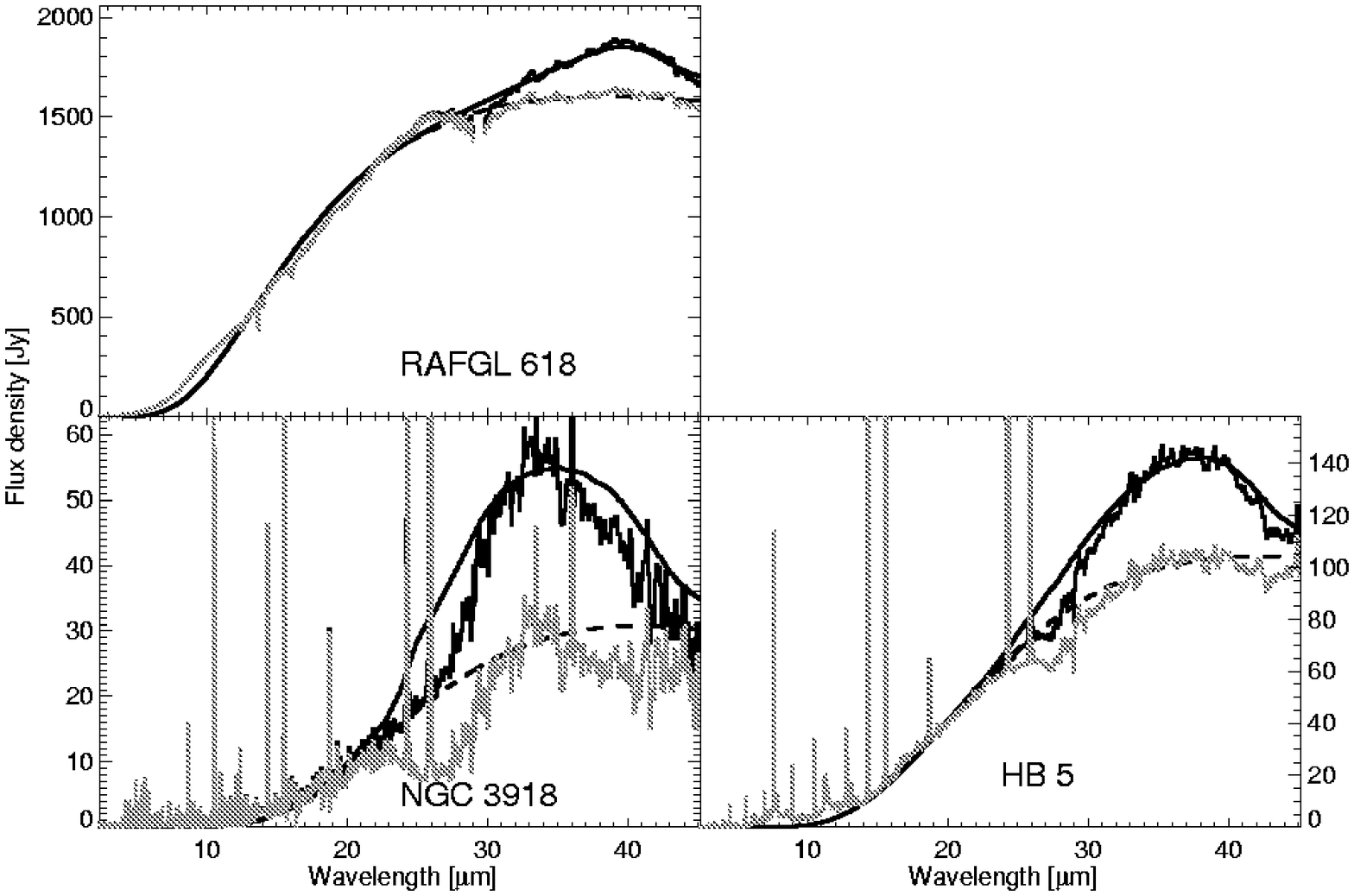}
  \caption{(continued).}
\end{figure*}
\begin{figure*}
  \includegraphics[width=17cm]{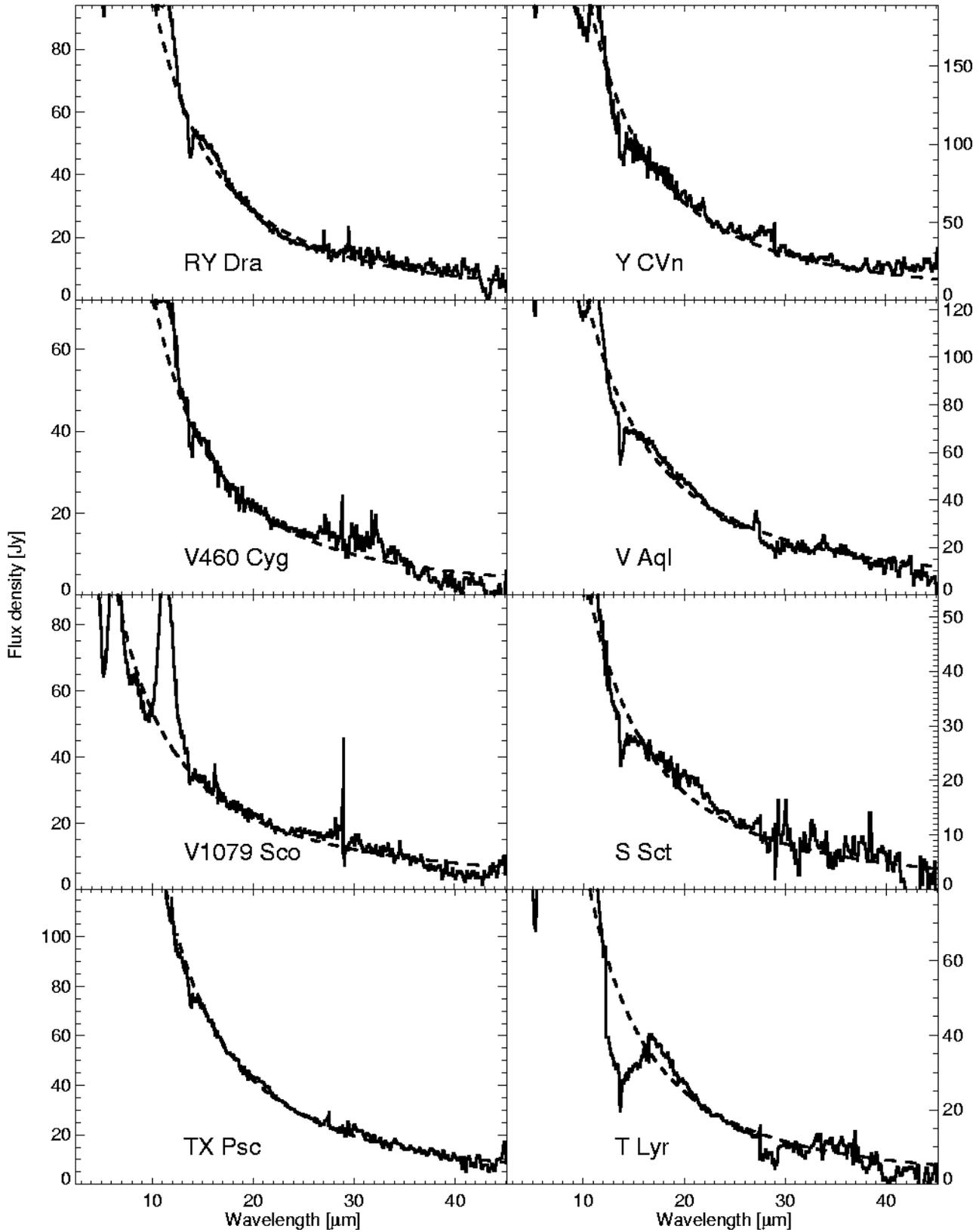}
  \caption{C-stars with a weak or without a ``30'' $\mu$m feature. We
    show the fitted continuum (dashed line). The sources are sorted
    according to decreasing continuum temperature. We refrain from
    showing a MgS model. In those cases where the strength of the
    ``30'' $\mu$m feature allowed to extract a temperature of the MgS
    the value is listed in Table~\ref{tab:properties}.}
  \label{fig:nofits}
\end{figure*}
\addtocounter{figure}{-1}
\begin{figure*}
  \includegraphics[width=17cm]{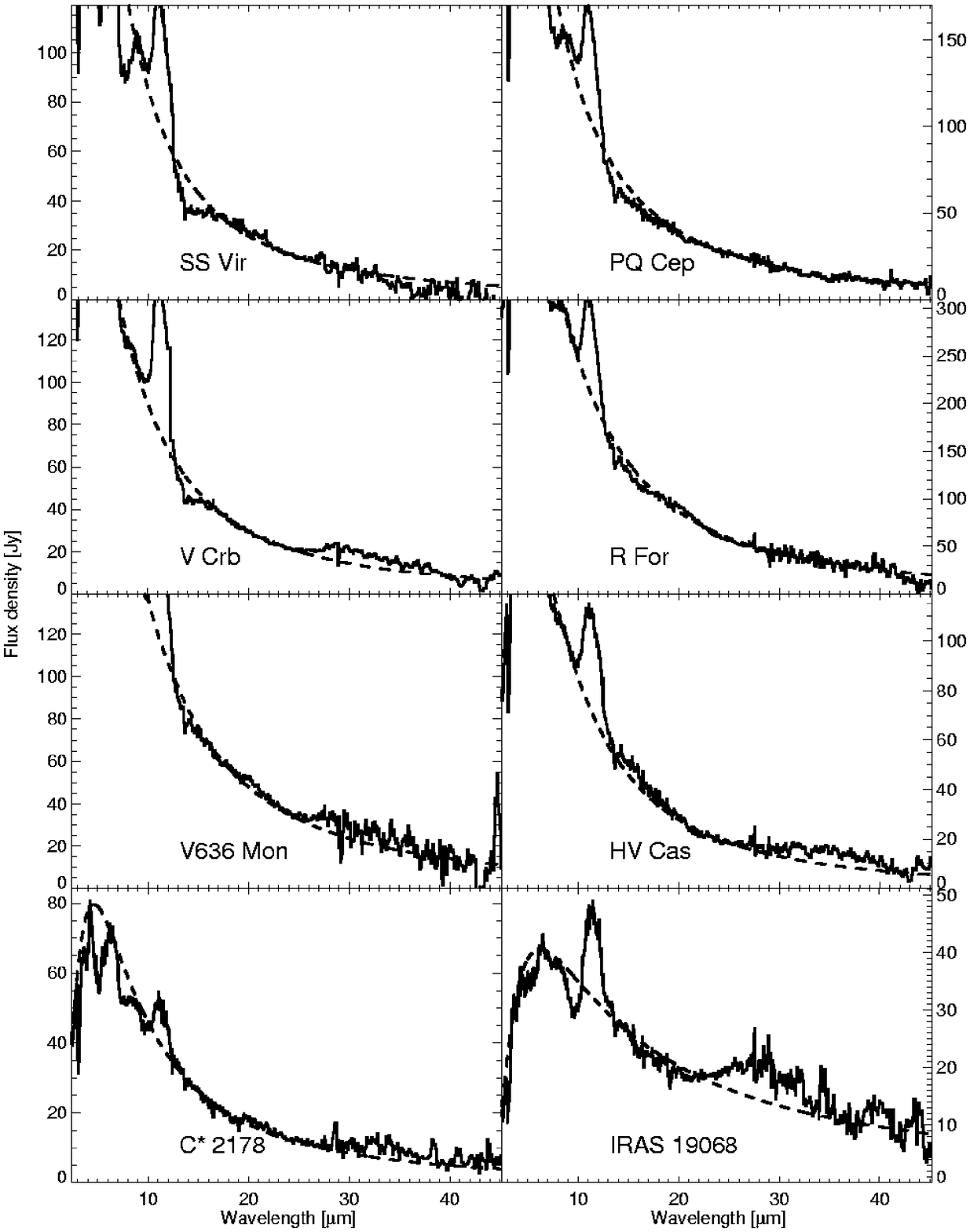}
  \caption{(continued).}
\end{figure*}
\end{document}